# Temperature-Dependent Diffuse Reflectance Measurements of Ceramic Powders in the Near- and Mid-Infrared Spectra

J. Michael Mayer[a], James A. Abraham[a], Bryan Kinzer[a], and Rohini Bala Chandran[a,*]

[a]*2350 Hayward St., G. G. Brown Building, Department of Mechanical Engineering, University of Michigan, Ann Arbor, MI 48109*

*Corresponding author. Email: rbchan@umich.edu, Phone: 734-647-9049

## Abstract

Radiative properties are critical to quantify radiative energy fluxes between surfaces and in participating media. However, there is limited experimental data on temperature-dependent radiative properties of materials. This work focuses on experimentally measuring temperature-dependent diffuse reflectance in the near- and mid-infrared spectra (1–20 μm) for ceramic particles with applications as heat-transfer and thermal-storage media in concentrated solar power (CSP) plants. Specifically, a commercially available sintered bauxite ceramic powder, ACCUCAST ID80, and its primary chemical constituents—alumina ($Al_2O_3$) and silica ($SiO_2$)—are measured using a Fourier transform infrared spectrometer (FTIR) coupled with a specialized diffuse reflectance accessory and a heated stage. Room-temperature diffuse reflectance measurements show increased absorption in tests with greater mass fractions of the ceramic samples. There is a strong correlation in the measured reflectance spectra of ACCUCAST with alumina and silica in the spectral range 2000–500 $cm^{-1}$ (5–20 μm). Whereas, for shorter wavelengths (<5 μm), the absorptance for ACCUCAST is greater than the absorptance for alumina and silica, indicating contributions from other chemical species present in the composite material. Transformations of the reflectance data with the Kubelka–Munk function show proportionality of absorption with mass fraction at selected wavelengths, but this pattern breaks down when sample reflectance approaches negligibly small values. For the first time, temperature-dependent diffuse reflectance measurements are reported for ACCUCAST, including a novel technique for accessing reflectance values above the limiting temperature of the background material KBr. These results are compared against those of silica and alumina through the calculation of a thermal emittance. All three materials exhibit a calculated emittance of ~0.9 at room temperature. However, this value drops to 0.68 at 1000 °C for ACCUCAST and ~0.43 for alumina and silica. Thermal cycling in air from 25 °C to 1000 °C resulted in a visible color change from dark grey to light orange for ACCUCAST and a subsequent 5X greater increase in reflectance at 4000 $cm^{-1}$ (2.5 μm) as compared to ACCUCAST thermally cycled at 1000 °C in vacuum. Alumina and silica spectra proved to be largely unaffected by thermal cycling under atmospheric and evacuated conditions. Overall, this study establishes a powerful technique for the characterization of radiative properties of particulate materials as a function of temperature and presents a detailed case study of ACCUCAST, a candidate for next-generation particle-based CSP.

# Nomenclature

| | |
|---|---|
| $E$ | emissive power |
| $d$ | diameter |
| $h$ | height |
| $I$ | intensity |
| $k$ | Kubelka–Munk absorption coefficient |
| $n$ | number of trials |
| $R$ | reflectance |
| $s$ | Kubelka–Munk scattering coefficient |
| $T$ | temperature |

*Greek letters*

| | |
|---|---|
| $\alpha$ | absorptance |
| $\delta$ | change in quantity |
| $\varepsilon$ | emittance |
| $\kappa$ | absorption coefficient |
| $\lambda$ | wavelength |
| $\mu$ | mean value |
| $\sigma$ | scattering coefficient |
| $\omega$ | scattering albedo |

*Subscripts*

| | |
|---|---|
| b | blackbody |
| exp | experimental |
| th | thermal |
| $\lambda$ | spectral |
| $\infty$ | relating to a semi-infinite medium |

*Superscripts*

| | |
|---|---|
| $'$ | directional |
| $\cap$ | hemispherical |

*Abbreviations*

| | |
|---|---|
| CSP | concentrated solar power |
| DRS | diffuse reflectance spectroscopy |
| DTGS | deuterated triglycine sulfate |
| FTIR | Fourier transform infrared spectrometer |
| K–M | Kubelka–Munk |
| PXRD | powder x-ray diffraction |
| SNR | signal-to-noise ratio |
| TGA | thermogravimetric analysis |
| UV–Vis | ultraviolet–visible spectroscopy |
| YSZ | yttria-stabilized zirconia |

# 1. Introduction

Radiative energy transport in particulate media plays a crucial role in many applications including receivers/reactors for concentrated solar power [1–5], photocatalysis with semiconductor particles [6,7], and powder bed additive manufacturing [8,9]. In some of these applications, the materials can experience elevated temperatures (~1000 °C in a solar receiver, for example), and the contributions of thermal radiation to overall heat transfer become significant. A quantitative heat transfer analysis of such a system requires the computation of radiative energy fluxes between surfaces and in participating media wherever applicable [10,11]. For opaque materials and surfaces, radiative properties such as emittance, absorptance, and reflectance are defining parameters that govern radiative energy exchange. In participating media, volumetric radiative properties—extinction and scattering coefficients and the scattering phase function—become pertinent. This study focuses on performing experimental measurements to determine spectrally resolved and temperature-dependent radiative properties for ceramic materials in particulate form.

At temperatures exceeding 1000 °C, many ceramic materials do not incur phase change and can exhibit excellent stability even in oxidizing and reactive environments. Therefore, they have broad applications as thermal barrier coatings [12,13], insulation and reactive media for concentrated solar receivers/reactors [14–17], reticulate foams for heat exchangers [18–20], and also heat-transfer fluids for concentrated solar power (CSP) technologies [21–23]. For example, Sandia National Laboratories (Albuquerque, USA) has developed a falling-particle solar receiver coupled with a moving packed-bed heat exchanger [23,24]. In this system, solid ceramic particles falling through a centralized tower receiver are heated directly by concentrated solar radiation. As a heat-transfer and energy storage medium, ceramic particles can enable power cycles with operating temperatures exceeding 700 °C, like the supercritical-$CO_2$ Brayton cycle [25], and provide round-the-clock electricity generation with sensible thermal energy storage [26,27]. Ideal particles for a solar receiver will have high absorptance in the incident solar spectrum (0.3–2.5 μm) and low thermal emittance in the re-radiation spectrum (1–20 μm) to capture and retain as much solar energy as possible [28–30]. Therefore, there is a necessity for spectral resolution in the measured radiative property data that spans the infrared spectrum to better quantify the overall heat-transfer behavior of materials at elevated temperatures.

Radiative properties of metals, non-metals, and coating materials were comprehensively compiled by Touloukian and DeWitt [31–33]. Additionally, radiative property data for man-made and natural materials are publicly accessible in online databases including the ECOSTRESS spectral library [34,35] and the Arizona State University spectral library for minerals, rocks, and soils [36]. While these databases are extensive in terms of compiling data from a wide variety of measurements, there are gaps in the reported data especially for temperature-dependent radiative properties. This issue is more pressing for less common and composite ceramic materials as compared to well-known materials such as alumina and silica. Because these data draw from wide-ranging experiments, there is often a sizeable spread in the reported data for the same materials, which makes it challenging to determine actual material properties. For example, normal spectral reflectance for alumina is reported at room temperature over the ultraviolet (UV), visible (VIS), and infrared (IR) spectra (0.23–37 μm) in ref. [32]. Within this dataset, measurements were made for different morphologies—powders, films, and crystals, different sample preparations—sintering, compacting, and polishing, and different reference standards—MgO, gold, or unspecified. Combined, these differences produce large disparities in the reported reflectance for even the same material. These factors can limit the usability of the data for quantitative studies, which is further exacerbated by the lack of digitization of many existing datasets.

Radiative properties of ceramics in powder form have been measured extensively at room temperature for various applications including dye-sensitized solar cells [37,38], aerogels for high-temperature insulation [39,40], and spectrally selective coatings for solar applications [41,42]. For concentrated solar and thermal energy storage applications, there is significant prior work on radiative property measurements of composite ceramic materials derived from sintered bauxite [29,43–47]. Bauxite particles used as hydraulic fracturing proppants and casting media are commercially available from CARBO Ceramics, Inc. (product names include CARBOHSP and ACCUCAST) [48,49]. Early studies looked at the scattering/extinction characteristics, and temperature-dependent diffuse reflectance of Masterbeads (Norton Chemical Company), a sintered bauxite proppant no longer in production [43,44]. Diffuse reflectance of the particles was measured from UV to near-IR, 0.3–2.5 μm, and it was found that a weighted solar reflectance varied by less than 1% for temperatures ranging from 150 °C to 930 °C. However, prolonged exposure to high temperatures (3 hours at 1000 °C) did increase reflectance by ~10% for most of the solar spectrum, indicating a need to evaluate the effects of thermal cycling on radiative material properties. More recently, Siegel et al. have measured spectral hemispherical absorptance (0.3–2 μm) and hemispherical emittance for packed beds of CARBO materials [29]. A weighted solar absorptance above 0.9 was reported for the particles and a thermal emittance of ~0.8 was reported at 700 °C for all bauxite particles measured; however, emittance is calculated from room temperature reflectance measurements made with an emissometer and by assuming that the particle emittance does not change with temperature. Ho et al. obtained solar absorptance and thermal emittance for sintered bauxite and sand particles [45]. Thermal cycling by exposing the particles at 1000 °C for 192 hours and cooling them back to room temperature was shown to reduce solar absorptance of the bauxite particles from 0.93 to 0.84. While the effects of exposure to high temperatures on the radiative behavior of the particles has been captured in this study, the properties were still measured at room temperature. Chen et al. have reported spectral directional-hemispherical reflectance at room temperature for sintered bauxite (CARBOBEAD) particles [46]. The authors combined experimental measurements with model-based interpretations to estimate the refractive index and extinction coefficient of these particles. These measurements are robust in terms of spectral resolution (0.38–15 μm), but the effects of temperature and thermal cycling on radiative properties are not characterized. Nie et al. performed comprehensive measurements for spectral directional-hemispherical absorptance (0.29–2.5 μm) and normal emittance (1–25 μm) for a bauxite-cordierite ceramic and several other ceramic powders [47]. Temperature-dependent normal emittance measurements of the bauxite ceramic showed that it decreases with temperature for 5–24 μm. From 600 °C to 1000 °C, normal emittance decreased by a maximum factor of ~10% over this spectral range. These results reinforce the necessity for improved quantification of temperature-dependent radiative material properties. Overall, there are substantial knowledge gaps on the effects of temperature and thermal history on the radiative behavior of ceramic powders for high-temperature applications. For composite ceramics derived from bauxite, there is also a lack of quantitative comparisons of the measured properties with its primary constituent oxides.

Motivated by the current knowledge gaps, the primary objective of this study is to experimentally determine spectral (1–20 μm) and temperature-dependent (25–1000 °C) diffuse reflectance behavior of ACCUCAST ID80, a commercially available sintered bauxite powder and candidate particle for CSP applications. These measurements are compared to and interpreted along with measurements of its dominant constituents, alumina ($Al_2O_3$) and silica ($SiO_2$).

Specifically, a Fourier transform infrared spectrometer (FTIR) coupled with a specialized diffuse reflectance accessory with a heated stage is used to collect radiative property data as a function of temperature. Additionally, the effects of thermal cycling at 500 °C and 1000 °C for 6 hours in both air and vacuum are investigated. Compared to prior work with similar scope, this study reports on advancements to: (1) quantify changes in reflectance and absorptance as a function of temperature through a combination of lower-temperature diluted sample measurements and high-temperature pure sample measurements, and (2) isolate the effects of the heating environment on radiative properties including ambient air and vacuum. The experimental methodology and approach presented in this work can be generalized and applied to evaluate radiative properties of a wide selection of materials and participating media.

## 2. Experimental Methods

### 2.1 Materials, Sample Preparation and Characterization

The three materials tested—alumina, silica, and ACCUCAST ID80—were all acquired in powder form directly, with the alumina and silica acquired in ultra-high purity forms due to the sensitivity of diffuse reflectance spectroscopy (DRS) to contaminants (Table 1).

*Table 1*. Summary of powdered materials used for testing, where average size is reported by the vendor for the material in its original, or "as received", state

| Material | Average Size | Vendor |
| --- | --- | --- |
| ACCUCAST ID80 | ~200 μm | CARBO Ceramics, Inc. |
| Alumina ($Al_2O_3$) | 3–5 μm | Inframat Advanced Materials |
| Silica ($SiO_2$) | 1 μm | US Research Nanomaterials, Inc. |

Particle size is an important factor in diffuse reflectance measurements. For reproducibility, it is recommended that the particles have an average size less than 25 μm [50]. It has also been observed that when volumetric scattering is the dominant scattering mechanism (over surface scattering), reflectance decreases with increasing particle size, in some cases significantly [51].

Consequently, the "as received" ACCUCAST particles were grinded using an yttria-stabilized zirconia (YSZ) mortar and pestle, and the size was controlled by running the ground powder through a 20 μm sieve. Particle size measurements were performed with a Beckman Coulter LS 13 320 particle size analyzer. The sieved powder was measured to have an average particle size of 12.4 μm with a standard deviation of 10 μm. Powder x-ray diffraction (PXRD) tests pre- and post-sieving confirm no observable change in chemical composition of ACCUCAST due to this process (Figure A.1). Optical images of the powders were acquired with an Olympus BX53M microscope coupled with an LC30 digital camera and an exposure time of 300 ms.

## 2.2 Diffuse Reflectance Spectroscopy

For materials in powder form, diffuse reflectance spectroscopy (DRS) presents a simple, yet sensitive way to probe volumetric absorption and scattering behavior of participating media [52–54]. Powdered samples can be readily tested with this approach with very minimal overhead. In this technique, powder sample materials are diluted and thoroughly mixed with a non-absorbing matrix material. For infrared measurements, potassium bromide (KBr) powder is used as a dilution agent due to its high transmittance for wavelengths up to 20 μm (>90% for 10 mm windows) [55]. This powder mixture is placed in a diffuse reflectance accessory within a spectrometer. Next, a source beam is directed at the mixture with concentrating optics. Part of the incoming radiation is specularly reflected at the surface of the powder bed and the remainder interacts with the bed volumetrically. This radiation is absorbed and scattered (a combination of reflection, refraction, and diffraction events) by the medium, which gives rise to diffusely scattered light out of the powder bed [56] (Figure 1b). The resulting intensity is collected and measured by the spectrometer and interpreted/reported as diffuse reflectance (Figure 1a).

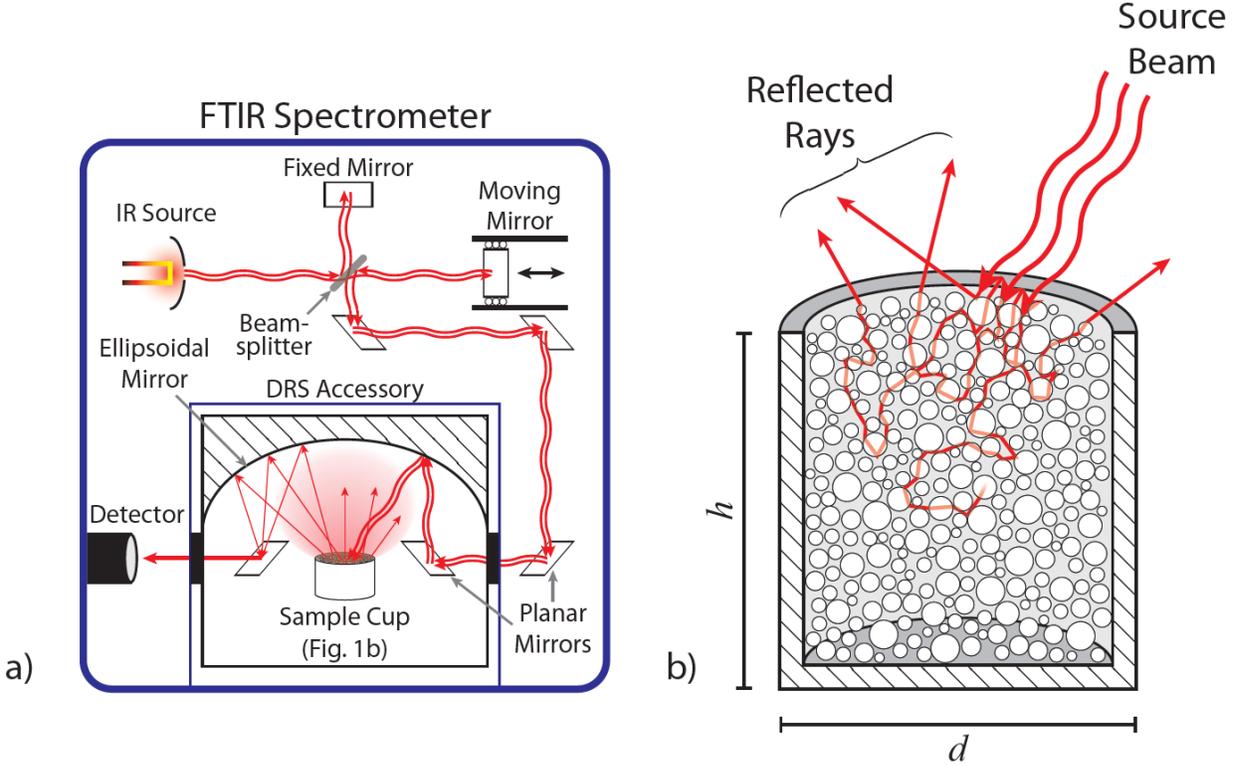

*Figure 1*. Schematics of (a) the measurement of a diffuse reflectance signal with a diffuse reflectance accessory and spectrometer housing the source beam and (b) diffuse reflection from the surface after scattering and absorption of the source beam by a bed of particles (represented by the red lines within the volume). Dimensions of the sample cup in (b) are $d = 10$ mm, $h = 2.3$ mm for the room-temperature accessory and $d = 5$ mm, $h = 2$ mm for the high-temperature accessory. Particles and sample cup not drawn to scale.

This measurement is done first with only the matrix material, KBr, which serves as a background against which the KBr + sample mixture is compared. Specifically, the diffuse reflectance, $R_{exp}$, at any wavelength $\lambda$ is the calculated as the ratio of measured intensity of the KBr + sample mixture to the measured intensity of the pure KBr matrix material (eq. 1).

$$R_{exp}(\lambda) = \frac{I_{KBr+sample}(\lambda)}{I_{KBr}(\lambda)} \tag{1}$$

DRS measurements are used to evaluate the effects of (1) sample mass fraction, (2) temperature, and (3) thermal cycling on the radiative properties for the materials listed in Table 1.

*Room-temperature DRS measurements:* In this study, a Thermo Fisher Scientific Nicolet iS50 FTIR was used with a PIKE Technologies DiffusIR diffuse reflectance accessory. The DiffusIR uses gold mirrors, including an ellipsoidal mirror, to guide the source beam from the FTIR into the powdered sample mixture with an angle of incidence of 30°. The IR source beam is produced by a silicon nitride filament heated up to 1300 °C (Polaris IR source, Thermo Fisher Scientific). To prevent unwanted absorption peaks in the measured spectra, the spectrometer and the DiffusIR are purged with air having moisture and $CO_2$ levels reduced to 1.5 and <1 ppm, respectively. The signal is measured with a DTGS (deuterated triglycine sulfate) detector over a wavelength range of 4000 $cm^{-1}$ to 400 $cm^{-1}$ (2.5–25 μm) at a resolution of 4 $cm^{-1}$, though data beyond 20 μm is noisy due to low signal throughput and therefore not used. To maximize the detected signal from the source beam, the slowest optical velocity of the moving mirror in the interferometer was chosen (0.1581 cm/s) and the aperture for the source beam was completely open. Additional measurements were performed with a tungsten-halogen white light source to extend the spectral range of measurements in the near-IR to 10,000 $cm^{-1}$ (1 μm). The white light source is used in the near-IR because of its higher intensity throughput compared to the IR source which drops off considerably in this range, leading to noisy data. Experimental parameters were kept the same for white light measurements except the data collection resolution was modified to 8 $cm^{-1}$ due to spectrometer limitations. Spectra were acquired and spectrometer settings were controlled with the software OMNIC from Thermo Fisher Scientific.

    To probe the effects of sample mass fraction on the measured diffuse reflectance, the powder samples were diluted with KBr in mass fractions ranging from 5% to 100% (pure sample). The mass fraction is defined as the ratio of the mass of the sample powder to the mass of the total mixture. To prepare the powders for measurement, they were thoroughly mixed with the matrix material KBr for ~3 minutes using a mortar and pestle. This step also ensures that any KBr crystals are ground to a fine powder. The sample mixtures fill a stainless-steel sample cup with diameter

10 mm and depth 2.3 mm, and the top of the sample cup is leveled using a razor, creating a flat and diffuse surface. A spectrum for the sample mixture is collected immediately following the collection of a background spectrum to eliminate the effect of any variables that may change with time, e.g., the composition of the purge environment.

The effects of thermal cycling on the measured diffuse reflectance properties at room temperature were also tested. This involved placing the ceramic powders in a tube furnace (Across International TF1700) at 500 °C or 1000 °C in an evacuated or air environment. The powders were heated to and held at this temperature for 6 hours and cooled back to room temperature. A maximum temperature of 1000 °C was selected based on the estimated upper bound for particles operating in a concentrated solar receiver, while 6 hours is representative of the minimum thermal storage timescales required in a concentrated solar power plant [2]. The heating rate was limited to 5 °C/min to ensure structural integrity of the alumina tube in the furnace.

*Temperature-dependent DRS Measurements:* High-temperature DRS measurements are possible using the same setup with a heated stage in place of a standard sample cup (Figure 2). The heated stage is composed of a ceramic heater connected to a power supply and temperature controller (PIKE Technologies) allowing it reach and hold temperatures as high as 1000 °C. The sample chamber is a small volume (~10 cm$^3$) surrounded by water-cooled steel walls and isolated from the rest of the accessory with an IR transparent window to preserve optical access. During measurements, this volume was evacuated with a vacuum pump (Agilent Technologies DS40M) to prevent the window from heating, heat loss to the environment, and potential reactions with oxygen in the air at elevated temperatures.

The temperature of the heated stage is measured with a type-K thermocouple and the accuracy is given as ±0.5% of the set point [57]. The temperature profile was programmed with the software TempPRO 7 from PIKE Technologies with a heating rate of 20 °C/min and hold times of up to 20 minutes to allow the sample temperature to stabilize within 0.1 °C of the set point temperature before a measurement is made. The reflectance was measured as the samples cooled from the maximum set point temperature, 500 °C or 1000 °C, to avoid the inclusion of volatile substances (e.g., adsorbed water) that are removed as the temperature of the sample increases during the initial heating cycle. Reflectance measurements of the same sample obtained during cooling and in the subsequent heating cycle vary by less than 3% at all points in the measured spectrum (Figure A.2).

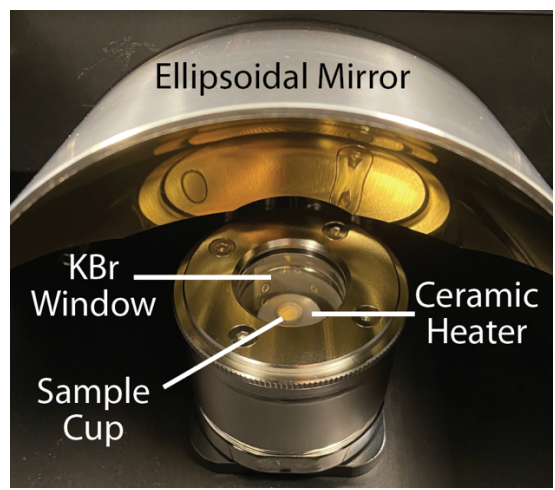

*Figure 2*. Picture of the heated stage inside the diffuse reflectance accessory showing the sample cup sitting in the ceramic heater. Sample cup dimensions are $d = 5$ mm, $h = 2$ mm for the high-temperature accessory. A window made of KBr or ZnSe is placed above the assembly which lies at the focus of the ellipsoidal mirror.

At 25 °C, reflectance measurements performed with the heated stage match well with measurements made using the room-temperature accessory (Figure A.3). Factors that contribute to deviations in reflectance stem from differences in using a different sample cup, which has a unique packing arrangement for the same mass fraction, and the use of a window in the high-temperature accessory. The window and smoother surface of the sample mixture can contribute to specular reflections which ultimately result in the reflectance spectrum increasing marginally as compared to the room-temperature accessory.

For high-temperature measurements, a smaller sample cup made of porous alumina (diameter of 5 mm and depth of 2 mm) is housed inside the ceramic heater (Figure 2). A press stick is used to pack the powder inside the sample cup to create a smooth, diffuse surface and to facilitate a uniform temperature due to improved thermal conductivity in the sample. A mass fraction of 5% was used for all tests in which the sample was diluted with KBr for temperature-dependent measurements. Due to its standard melting point of 730 °C and non-negligible vapor pressure at elevated temperatures, KBr was used as a matrix material for temperature-dependent reflectance measurements only up to 500 °C [58,59]. However, reflectance measurements with the pure samples (sample powders without the KBr matrix) were performed up to 1000 °C. IR-transparent windows made of ZnSe and KBr were used in these tests (thickness 3 mm, acquired from Knight Optical and Sigma Aldrich, respectively). Following tests in which the KBr was present in the sample cup, it was observed that a thin layer of KBr was deposited on the window surface. This is possibly explained by volatilization of the KBr due to the combination of high temperatures and low pressures. From 364 °C to 627 °C, the vapor pressure of KBr increases from 34 µPa to 3.35 Pa [59]. Thus, as temperature increases, the vapor pressure of KBr exceeds the rated pressure of the vacuum pump used to evacuate the sample chamber, 0.67 Pa [60]. Therefore, the ZnSe window was preferentially used for the tests involving KBr because the deposited KBr layer can be removed with distilled water, unlike with the KBr window [61]. For the pure sample tests up to 1000 °C, the KBr window was used because of its higher transmittance compared to ZnSe in the infrared spectrum [32]. Based on thermogravimetric analysis (TGA) measurements, the mass change of KBr as it is heated up to 500 °C is negligible, i.e., less than 0.05% of the starting mass. Therefore, this effect is not expected to have a measurable effect on temperature-dependent reflectance measurements. TGA measurements were performed with a TGA 550 from TA

Instruments at a pressure of 1 atm in an inert argon environment, however, and may not fully capture the loss of KBr due to sublimation in an evacuated environment.

Since a reflectance cannot be calculated without a KBr background measurement, a *relative* reflectance is calculated for tests up to 1000 °C involving the pure samples (eq. 2).

$$R_{\text{relative}}(\lambda, T) = \frac{I_{\text{sample}}(\lambda, T)}{I_{\text{sample}}(\lambda, 25\ °C)} \tag{2}$$

The relative reflectance of a material at a temperature $T$ is defined as the ratio of the measured radiative intensity of the pure sample, $I_{\text{sample}}(\lambda, T)$, at $T$ to the intensity measured at room temperature, $I_{\text{sample}}(\lambda, 25\ °C)$. While this ratio is not a direct measurement of the material's reflectance as a function of temperature, it does provide a quantitative understanding of how reflectance will change up to 1000 °C. This data for the pure sample measurements up to 1000 °C can be applied as a scaling factor to the corresponding room-temperature reflectance data to deduce reflectance at elevated temperatures (eq. 3). This results in extrapolated values of reflectance up to 1000 °C which normally cannot be obtained for measurements that utilize KBr as a background.

$$R_{\text{extrapolated}}(\lambda, T) = \frac{I_{\text{sample}}(\lambda, T)}{I_{\text{sample}}(\lambda, 25\ °C)} R_{\text{exp}}(\lambda, 25\ °C) \tag{3}$$

Pure sample reflectance measurements are shown alongside extrapolated reflectance values at the same temperature up to 500 °C in Figure A.4. Contributions of emission to the overall measured signal are neglected when calculating reflectance at elevated temperatures. This is a reasonable assumption because the DRS accessory is not optimized to collect emission signals, and the measured spectra are dominated by the signal coming from the source beam and reflected by the sample. Figure A.5 shows that the emitted intensity is a small fraction of the measured intensity even at 1000 °C. Moreover, because the spectra acquired with the source beam turned off can be impacted by various other components (e.g., heated surfaces) in addition to the heated sample, it is not tractable to only account for emission from the sample.

*UV–Vis DRS Measurements:* Diffuse reflectance measurements were also performed on a UV–Vis spectrometer (Shimadzu UV-2600) with an integrating sphere (ISR-2600) for ACCUCAST. This was necessitated by observations of visible color change when the ACCUCAST powder was subjected to elevated temperatures in both the tube furnace and the heated stage when air was present. These measurements used barium sulfate ($BaSO_4$) as a matrix/background material [62]. ACCUCAST used in the UV–Vis measurements was diluted to a mass fraction of 10%.

**2.3 Data Analyses and Experimental Repeatability**

*Kubelka–Munk Transform:* The Kubelka–Munk (K–M) transform was applied for select spectroscopic measurements to further understand the relationship between absorption and scattering behavior in sample mixtures as a function of its composition. Kubelka–Munk theory can reasonably approximate radiative transport in a homogeneous, plane-parallel (i.e., one-dimensional), and isotropically scattering medium. By additionally assuming radiative intensity to be isotropic over the incident and reflected hemispheres, the radiative intensity field can be described by two components—one along the direction of incident flux and one in the opposite direction. Importantly, this set of assumptions reduce the complex differential-integral radiative transport equation into a pair of ordinary differential equations with closed-form analytical solutions, which have been extensively discussed in refs. [63–66]. For an optically thick, semi-infinite medium, the reflectance at the surface, which is equivalent to an experimental diffuse reflectance, is calculated as (eq. 4),

$$R_\infty = 1 + \frac{k}{s} - \sqrt{\frac{k}{s}\left(2 + \frac{k}{s}\right)} \qquad (4)$$

where $k$ and $s$ are related to the absorption and scattering coefficients, $\kappa$ and $\sigma$, respectively, of the participating medium (eqs. 5a–b).

$$k = 2\kappa \tag{5a}$$

$$s = \sigma \tag{5b}$$

The Kubelka–Munk function, $f(R_\infty)$, compares the square of the absorptance of the medium to the surface reflectance and can be succinctly expressed as the ratio $k/s$ as shown in eq. (6) and is plotted as a function of $R_\infty$ in Figure A.6.

$$f(R_\infty) = \frac{(1 - R_\infty)^2}{2R_\infty} = \frac{k}{s} \tag{6}$$

$$\frac{k}{s} = 2\left(\frac{1}{\omega} - 1\right) \tag{7}$$

The ratio $k/s$ can be related to the scattering albedo, $\omega$, which dictates the probability of scattering in attenuation events within a radiatively participating medium (eq. 7). The $k/s$ ratio and therefore the K–M function increase as the value of $\omega$ decreases, or equivalently when the medium becomes relatively more absorbing than scattering. Hence, the K–M function (eq. 6) can be directly correlated to the absorption behavior of the sample material present in the highly scattering KBr matrix in the powders tested. Additionally, the K–M transform strongly emphasizes regions of high absorption in the measured spectra and is therefore useful for identifying absorption peaks.

*Spectrally-Weighted Quantities*: Thermal emittance was calculated using the measured reflectance spectra. The DRS accessory measures the spectral, directional-hemispherical diffuse reflectance from the surface of the powdered samples. A spectral, directional absorptance can be obtained from the measured reflectance from eq. (8), which assumes zero transmittance from the powders present in the sample cup. This assumption can be justified with an estimated wavelength-averaged scattering coefficient of 134 mm$^{-1}$ for KBr powder having a uniform particle size of 10 µm and a volume fraction of 0.36. These estimates lead to a large optical thickness of the medium which precludes any transmitted radiation from leaving the sample cup. Spectral refractive indices of KBr at room temperature were calculated using a dispersion equation [67], and the absorption coefficient was assumed to be negligible over the measured wavelengths. The addition of sample materials is expected to increase extinction in the medium because similar particle sizes will result in comparable scattering, but absorption will no longer be negligible. Therefore, the assumption of zero transmittance is expected to hold for all measurements regardless of sample concentration. Consequently, eq. (8) can be applied to relate spectral directional absorptance, $\alpha'_\lambda$, with the spectral directional-hemispherical reflectance, $R'^{\cap}_\lambda$, and the measured spectral diffuse reflectance, $R_{\exp}(\lambda, T)$.

$$\alpha'_\lambda(T) = 1 - R'^{\cap}_\lambda(T) = 1 - R_{\exp}(\lambda, T) \tag{8}$$

$$\alpha_\lambda(T) = \alpha'_\lambda(T) \tag{9}$$

$$\alpha_\lambda(T) = \varepsilon_\lambda(T) \tag{10}$$

Because the samples measured are ensembles of particles with rough surfaces, it is reasonable to assume a diffuse absorptance behavior, which results in the equality of the spectral directional absorptance and the spectral hemispherical absorptance in eq. (9). Applying Kirchhoff's law for local thermodynamic equilibrium, it can be deduced that this spectral hemispherical absorptance will be equal to the spectral hemispherical emittance (eq. 10) of the particulate medium [68]. Thus, a hemispherical thermal emittance for the particle ensemble can be calculated (eq. 11) using the experimental reflectance data collected from 1 µm to 20 µm.

$$\varepsilon_{\text{th}}(T) = \frac{\int_{1\ \mu m}^{20\ \mu m}[1 - R_{\exp}(\lambda, T)]E_{b\lambda}(T)\,d\lambda}{\int_{1\ \mu m}^{20\ \mu m} E_{b\lambda}(T)\,d\lambda} \tag{11}$$

The thermal emittance can be calculated precisely with knowledge of the material properties over a spectral range that intersects with a majority of the blackbody emission spectrum. In this study, the spectral range is limited to 10000–500 cm$^{-1}$ (1–20 μm) which is dictated by the source/beamsplitter/detector combination used in the measurements. This spectral range accounts for >95% of the blackbody emissive power for temperatures ≥400 °C and ~99% at 1000 °C. As temperature decreases, eq. (11) deviates more from a true thermal emittance. For example, the spectral range 1–20 μm accounts for less than 74% of blackbody emissive power at 25 °C. However, eq. (11) can still be used to identify trends in thermal emittance with temperature.

In applying eq. (11), reflectance values obtained for the pure samples (and not a sample mixture with KBr) were utilized to calculate emittance. Extrapolated reflectance data was acquired using eq. (3) to get values up to 1000 °C. Reflectance values that exceeded 100% were considered the same as reflectance values of 100% in this calculation due to the unphysical nature of having a negative absorptance/emittance. Data from the IR source was used in the spectral range 2.5–20 μm, while data from the white light source was used from 1 μm to 2.5 μm. To impose continuity in reflectance, the white light source data is offset to match the IR source data at 2.5 μm. This approach is adopted because the spectrometer components and reflectance accessory are optimized for the near- and mid-infrared wavelengths and to maintain consistency with all other data presented in this study, measured using the IR source. The maximum difference in thermal emittance calculated by imposing continuity is less than 1.6% of the emittance calculated without the offset for all materials and temperatures.

*Data Repeatability and Error Analyses*: Systematic/instrumental uncertainties and specifications for relevant measured quantities are summarized in Table 2. Although different sources of random error could be present in the measurements, the main source of variation is attributed to sample preparation. Preparing the sample requires a series of steps that can be challenging to control precisely, and this factor has been observed to strongly influence the measured reflectance spectra. For example, there can be minor variations in the packing density, particle size distribution, and the orientation and morphology of particles on the surface due to sample preparation and loading. Thus, repeatability is verified in these measurements by inspecting deviation from the mean in the spectral reflectance measured for three tests having the same conditions, where the sample mixture and background have been prepared from scratch for each test. Experimental error was quantified by averaging the deviations in spectral reflectance, $R_\lambda$, from the mean values, $\mu_\lambda$, calculated from three experimental trials, over all wavelengths recorded for the measurement (eq. 12).

*Table 2.* Uncertainty and specifications of measured experimental parameters [57,85,86]

| Equipment | Parameter | Uncertainty/Specs |
|---|---|---|
| DTGS detector | SNR (Voltage) | >13,000:1 |
| FTIR | Wavenumber | ±0.01 cm$^{-1}$ |
| Furnace | Temperature | ±5 °C (maximum) |
| Heated stage | Temperature | ±5 °C (maximum) |
| Mass balance | Mass | ±0.0001 g |

$$\overline{\delta R} = \frac{1}{3n} \sum_{\text{Trial 1}}^{\text{Trial 3}} \sum_{\lambda=\lambda_1}^{\lambda_n} |R_\lambda - \mu_\lambda| \tag{12}$$

Moreover, a 95$^{\text{th}}$ percentile deviation from the mean was quantified to determine near-maximum deviation in reflectance without the skewed influence of outliers. This was measured by identifying the value below which 95% of deviations from the mean over all tests and wavelengths exist. To assess the effectiveness of performing three tests, a measurement with ten tests was performed for one sample (ACCUCAST, 5% mass fraction). Across all different combinations of three tests selected from these ten tests, i.e., $_{10}C_3$ or 120 combinations of datasets, the largest average deviation, $\overline{\delta R}$, was 2.04% and the smallest was 0.16%. Therefore, while three tests per measurement do not completely capture test-to-test variation in the reflectance measurements, the errors are still small enough to justify its use and decrease experimentation time.

Twenty scans were performed per sample to verify the spectrum was not changing with time and to reduce noise with averaging. The difference in intensity measured by the detector between 20 scans and 100 scans was calculated to be <1% at any wavelength, which suggests that a greater number of scans provide no additional accuracy.

## 3. Results and Discussion

### 3.1 Room-Temperature Spectroscopic Measurements

Figures 3a–c present the effects of mass fraction (5–100%) on room-temperature diffuse reflectance measurements for each ceramic powder listed in Table 1. Each spectrum in Figure 3 is the average over three trials. For mass fractions less than 5% (not shown), while the measured reflectance is larger, the spectral trends and features match those obtained for the larger mass fractions. Each line represents the mean value of three tests performed for each material.

The spectra for ACCUCAST (Figure 3a) show a steady increase in reflectance from 4000 cm$^{-1}$ (2.5 μm) to a peak around 2150 cm$^{-1}$ (4.65 μm) for all mass fractions. This is followed by a steep drop to a minimum reflectance (i.e., maximum absorptance) at 1150 cm$^{-1}$ (8.7 μm), which was also observed in ref. [46] for unaltered CARBOBEAD particles. Finally, this is followed by a small spike in reflectance near 1065 cm$^{-1}$ (9.4 μm) before the values decrease and level off. This reflectance spike is only observed up to mass fractions of 40% before absorption dominates the spectrum, which also matches with reported data for beds of bauxite particles tested in ref. [46]. The spectra for alumina (Figure 3b) start with reflectance values above 100% in the near-infrared for all mass fractions, which indicates that alumina is more reflective than the standard (KBr) up to 3850 cm$^{-1}$ (2.6 µm). The alumina spectra continue with a local minima in reflectance at 3475 cm$^{-1}$ (2.88 µm), indicative of O-H stretching vibrations and the presence of hydroxyl groups due to adsorption of water on the particle surfaces [69]. This dip is followed by an increase and leveling off to gray behavior from 2600 cm$^{-1}$ to 2000 cm$^{-1}$ (3.85–5 µm) in which the pure alumina exhibits a reflectance in the range 75–80%. The wide absorption band around 700 cm$^{-1}$ (14.3 µm) due to Al-O stretching vibrations is observed for all mass fractions of alumina and consistent with reported spectra for alumina [70]. A strong absorption peak at 1025 cm$^{-1}$ (9.76 µm) is only observed for the 100% mass fraction curve with a reflectance that is nearly zero and reported in ref. [71] for pure alumina powders and crystalline forms of alumina like the mineral corundum [34–36]. In the silica spectra (Figure 3c), similar trends as those seen in alumina are observed for 4000–2500 cm$^{-1}$ (2.5–4 µm), due to absorption associated with hydroxyl groups. Beyond 2300 cm$^{-1}$ (4.35 µm), oscillations are present in the spectral range 2180–1440 cm$^{-1}$ (4.6–6.9 µm), representing combination and overtone bands [72]. A strong absorption peak centered on 1100 cm$^{-1}$ (9.1 µm) aligns with Si-O-Si asymmetric stretching vibrations reported in ref. [73], while additional absorption bands out to 500 cm$^{-1}$ (20 µm) are attributed to Si-O-Si symmetric stretching vibrations, which correlate well with reported spectra in ref. [72].

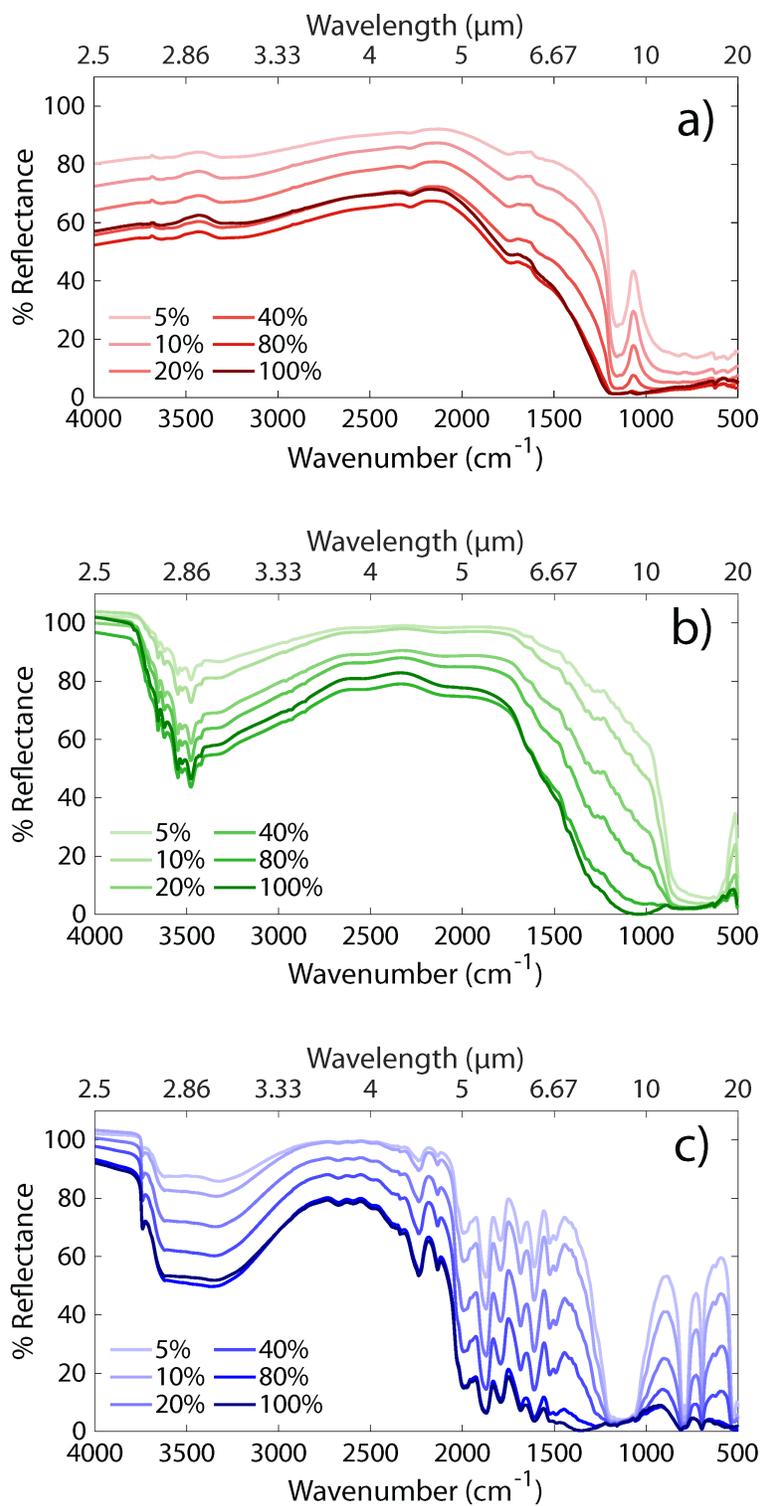

*Figure 3.* Room-temperature diffuse reflectance spectra averaged over three trials, for a) ACCUCAST ID80, b) alumina, and c) silica for mass fractions ranging from 5% to 100% for a spectral range of 4000–500 cm$^{-1}$ (2.5–20 μm). Error bars are not shown for clarity of the image, but is shown in Figure 4.

For all materials tested, there is a monotonic trend of reflectance decreasing with increasing mass fraction up to 80% due to increasing concentration of the absorbing species in the sample and KBr mixture. However, from a mass fraction of 80% to 100%, the margin of change in reflectance decreases, and is within the average deviations in the measured reflectance at these mass fractions. For example, the average reflectance value of ACCUCAST at 2000 cm$^{-1}$ (5 μm) decreases by 28.1% as the mass fraction increases from 5% to 80%. When the mass fraction increases to 100%, the reflectance increases by 1.3%, which is within experimentally measured sample-to-sample deviations of 2.64% and 2.23% for 80% and 100% mass fractions, respectively.

Figure 4 shows the average and 95$^{th}$ percentile deviations from the mean value due to sample-to-sample variations for each material as a function of mass fraction. Spectral dependence of deviation cannot be observed from this plot, but regions of high reflectance exhibit the largest deviation. Whereas the reflectance in these regions is subject to random differences across samples, the reflectance in regions of high absorption collapses to a small value, such as for the absorption peak seen in silica at 1100 cm$^{-1}$ (9.1 μm), little variation is observed. Figure 4 reveals that for all samples and mass fractions tested, average and 95$^{th}$ percentile deviations, from the mean reflectance reported in Figure 3, are less than 5% and indicates reasonable repeatability in the measurements. In general, these deviations increase for higher mass fractions, and this behavior is attributed to the increased influence of absorption and scattering events near the surface of the sample. With less KBr present, less of the incident radiation is volumetrically scattered into the mixture, and therefore the radiative behavior of the sample contributes less to the measured reflectance. Because the surface roughness is subject to random differences across each test, there is an increase in deviation in the measured data. From these results, it is concluded that the diffuse reflectance technique is more consistent and reproducible as the sample becomes more and more diluted. Moreover, dilution also increases the sensitivity of the measurement to spectral features, e.g. the local maximum in reflectance at 1065 cm$^{-1}$ (9.4 µm) for ACCUCAST (Figure 3a). Because the 5% mass fraction mixtures yielded the best combination of reproducibility (<2% average deviation for all samples tested) and sensitivity in the regions of high absorption, this mass fraction is down-selected for evaluating the effect of thermal cycling. Temperature-dependent diffuse reflectance measurements were performed on samples with mass fractions of 5% and 100%.

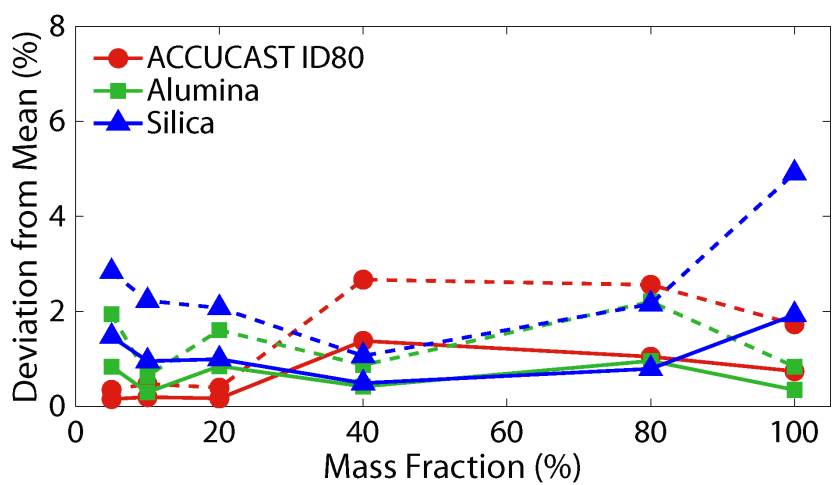

*Figure 4*. Average (solid lines) and 95$^{th}$ percentile (dashed lines) deviations from the mean reflectance (Figures 3a–c) for each material as a function of mass fraction, averaged over three tests and all wavelengths.

To further interpret the effects of mass fraction, Figures 5a–c transform the average reflectance spectra shown in Figures 3a–c to K–M units (eq. 6). Due to the rapid change in K–M values with increasing absorption (Figure A.6), Figures 5a–c are plotted with different scales on the x- and y-axes for each material. For all mass fractions, K–M transformation indicates that scattering of incident radiation is more dominant than absorption, i.e., $k/s < 1$, for wavenumbers >1500 cm$^{-1}$ (< 6.67 µm) for ACCUCAST and alumina, and >2500 cm$^{-1}$ (< 5 µm) for silica. Consequently, these spectral regions are not included in Figures 5a–c. For alumina and silica, K–M values tend toward infinity for wavelengths at which the reflectance is nearly zero, i.e., the material is completely absorbing and opaque at these wavelengths. This is consistent with eq. (6), where the denominator of the K–M transform becomes zero when reflectance becomes zero. For all materials, K–M values generally increase with mass fraction due to increased absorption in the KBr + sample mixture. This increase is approximately proportional to the increase in mass fraction for most wavelengths, but this behavior is not always observed in regions with high absorption, e.g., alumina for mass fractions ≥80%.

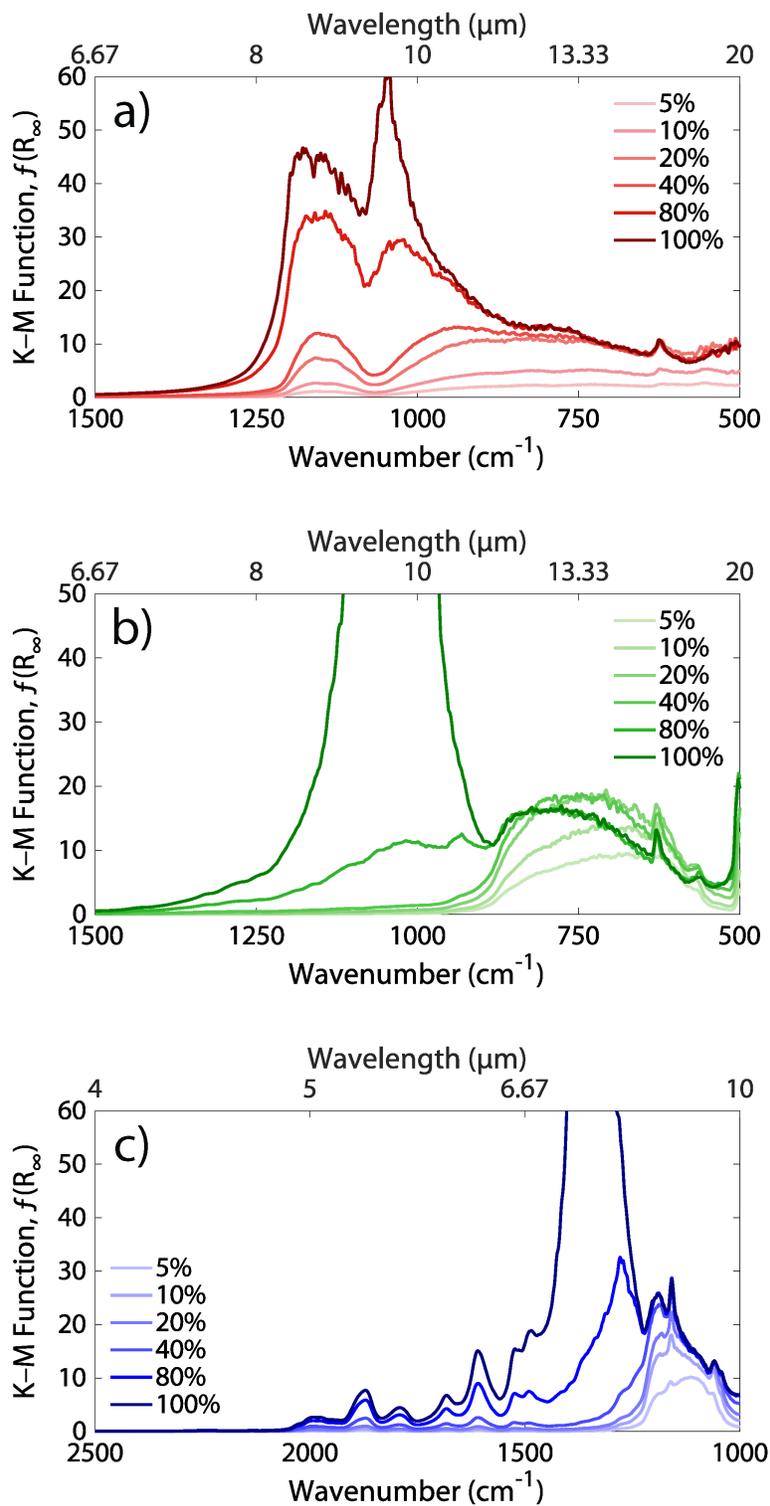

*Figure 5.* Room-temperature diffuse reflectance spectra after the Kubelka–Munk transform for a) ACCUCAST ID80, b) alumina, and c) silica for mass fractions ranging from 5% to 100% fraction at 1000 cm$^{-1}$ (1 μm) for each material.

Figure 6 shows the ratios of $f(R_\infty)$, the K–M-transformed reflectance values, at a given mass fraction with the K–M value at 5% mass fraction at selected wavelengths of absorption peaks. These ratios are compared to the ratios of the mass fractions with 5%, which follows a linear function. For perfectly proportional scaling, the ratios of $f(R_\infty)$ obtained at every mass fraction will exactly match the ratio of the mass fractions. For example, at 1150 cm$^{-1}$ (8.7 μm) the K–M values for ACCUCAST increase from 1.08 to 2.59 as the mass fraction increases from 5% to 10%. This gives a K–M ratio of 2.40 for a mass fraction ratio of 2, and this ratio is plotted at 10% mass fraction. The ratio of mass fractions is consistently doubled up to 80%, where the ratio of mass fractions is 16. At 100% mass fraction, the ratio reaches 20. These mass fraction ratios are included in Figure 6 as a benchmark against which the ratios of K–M units can be compared. For all three materials, the K–M values change more-or-less proportionally with the mass fraction up to 20%. As the mass fraction increases, the ratio of $f(R_\infty)$ deviates further from the mass fraction ratio. At 80% mass fraction, the ratio of $f(R_\infty)$ for ACCUCAST at 1150 cm$^{-1}$ (8.7 μm) is greater than the mass fraction ratio by a factor of 2. For the same mass fraction, silica differs by a factor of 3.6, and alumina differs by a factor 5.1. This factor increases again for all three materials at 100% mass fraction, and it is especially apparent for alumina which differs by a factor of 130 at the selected absorption peak.

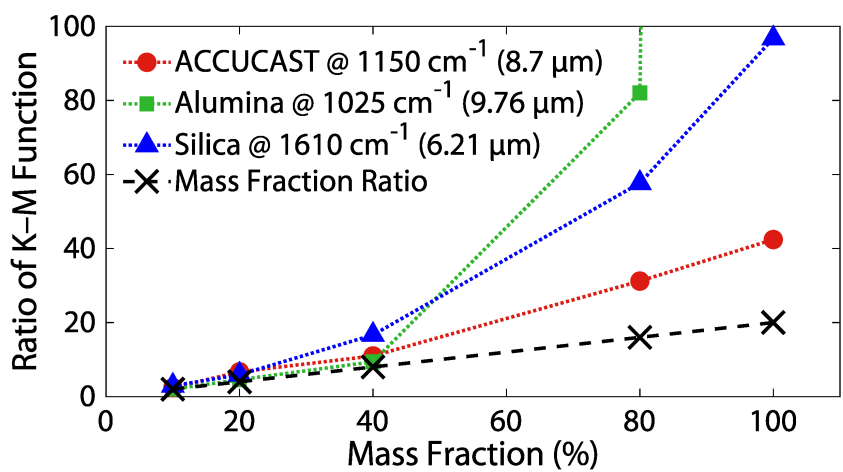

*Figure 6.* Ratios of K–M function, $f(R_\infty)$, at selected wavelengths for ACCUCAST (8.7 μm), alumina (9.76 μm), and silica (6.21 μm) as a function of mass fraction, compared to the true mass fraction ratios. Ratios are with respect to a mass fraction of 5%. Dashed lines are included to exhibit trends in the ratios as a function of mass fraction.

Proportional change in K–M values signifies that the absorption coefficient $k$ is changing proportionally with the concentration of the absorbing species and therefore its mass fraction, while the scattering coefficient $s$ remains largely unaltered. This behavior is expected for small mass fractions where scattering within the powder bed is mostly dominated by the IR-transparent KBr as compared to the sample material. At higher mass fractions, the extent of scattering from KBr decreases, and therefore both the absorption and scattering behavior of the medium start to change. As a result, the K–M values no longer exhibit strict proportionality with mass fraction, as seen with each material beyond 40%. Moreover, the inherent assumptions that relate the transformed reflectance to the ratio of absorption and scattering coefficients (eq. 6) become less valid as the mass fraction of absorbing species increases, leading to the incident radiation interacting mostly near the surface of the powder bed instead of its entire volume. This is especially true for wavelengths that are highly absorbing, such as in the case of alumina at 1025 cm$^{-1}$ (9.76 μm), when absorption is so dominant that nearly no radiation scatters out of the sample.

To understand the influences of the constituent oxides of ACCUCAST on its radiative properties, Figure 7a presents a comparison of the average reflectance spectrum for ACCUCAST with (a) alumina, (b) silica, and additionally (c) a mixture with a 5:1 mass ratio of alumina and silica powders. Each spectrum is obtained for a sample mass fraction of 5% (with the balance being KBr). Figure 7b shows the K-M transformed reflectance to highlight the comparison of absorption features. The alumina/silica binary mixture with a 5:1 mass ratio was chosen to approximate the composition of ACCUCAST, of which alumina and silica make up roughly 90% by mass [49], even if this excludes some key constituents such as iron oxide ($Fe_2O_3$) and titanium dioxide ($TiO_2$).

For alumina and silica (as well as the binary mixture), there is a dip in reflectance from 3750 to 2600 cm$^{-1}$ (2.67 to 3.85 μm). While this behavior is not observed for ACCUCAST, it exhibits a lower reflectance overall in the spectral range 4000–2000 cm$^{-1}$ (2.5–5 μm). This indicates that another species present in ACCUCAST, most likely $Fe_2O_3$ as it can absorb energy in these wavelengths [74,75]. In the wavenumber range 2000–1000 cm$^{-1}$ (5–10 μm), the ACCUCAST reflectance spectrum lies in between that obtained for alumina and silica, and closely matches the reflectance spectrum for the binary mixture. However, the oscillatory behavior observed in the silica spectrum over this range does not appear in the ACCUCAST spectrum. This is possibly due to the smaller proportion of silica in the overall makeup of ACCUCAST (given as 10–20%). In addition, the strong absorption peak at 1100 cm$^{-1}$ (9.1 µm) associated with silica is more pronounced (higher absorption) in the binary mixture versus the ACCUCAST (Figure 7b).

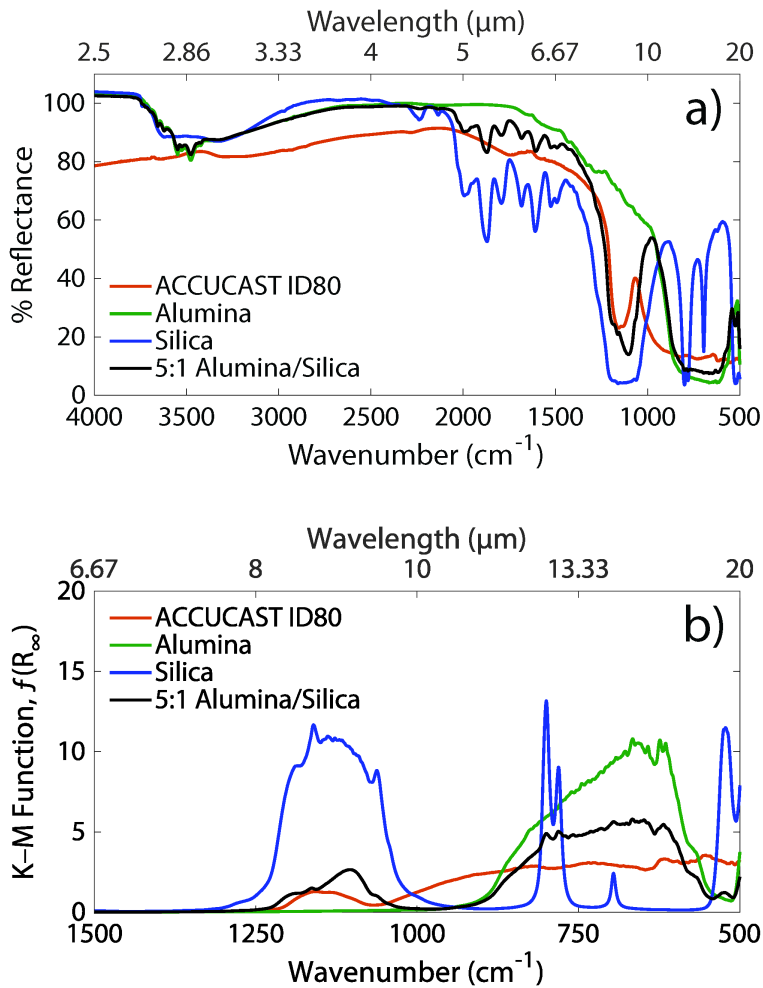

*Figure 7*. Room-temperature diffuse reflectance spectra a) for each material (including the 5:1 alumina/silica mixture) at 5% mass fraction and b) after the Kubelka–Munk transform for a spectral range of a) 4000–500 cm$^{-1}$ (2.5–20 μm) and b) 1500–500 cm$^{-1}$ (6.67–20 μm).

The strength of correlation between the various spectra shown in Figure 7a was quantified by calculating an average deviation in reflectance and a Spearman correlation coefficient over the full spectrum and for separate spectral bands. Based on average deviation, the ACCUCAST spectrum is most strongly correlated with the spectrum obtained for the binary mixture, followed by alumina and then silica. The average deviation in the reflectance spectra from the mean of the ACCUCAST spectrum is 9.3% for the binary mixture, 12.8% for alumina, and 15.5% for silica. In the spectral range 2000–500 cm$^{-1}$ (5–20 µm), the average deviation between the spectra obtained for ACCUCAST and the binary mixture becomes even smaller (7.9%), indicating that the radiative behavior of ACCUCAST in this part of the spectrum stems heavily from alumina and silica. This gives rise to a local maximum in reflectance between the main alumina and silica absorption bands, a feature that is observed in both the binary mixture and ACCUCAST. Figure 7b shows that the characteristic wavenumber/wavelength of this absorption peak for ACCUCAST, 1060 cm$^{-1}$ (9.4 µm), differs slightly from that of the absorption peaks of the binary mixture, 980 cm$^{-1}$ (10.2 µm), and silica spectrum, 1100 cm$^{-1}$ (9.1 µm). Based on the ratio of the K–M values at this wavelength, which roughly scales with mass fraction (Figure 6), it can be inferred that ACCUCAST has roughly half the concentration of silica and less alumina as that present in the binary mixture.

A Spearman correlation coefficient was additionally computed to quantify how well two spectra linearly correlate with each other [76]. For the 4000–500 cm$^{-1}$ (1–20 μm) spectrum, ACCUCAST is most strongly correlated with alumina (0.72), closely followed by the binary mixture (0.70), and then silica (0.62). In the spectral range 2000–500 cm$^{-1}$ (5–20 μm), the Spearman correlation coefficients increase for all three spectra—0.97 for alumina, 0.91 for the binary mixture, and 0.67 for silica. Thus, both correlation metrics are indicative that the radiative properties of ACCUCAST are strongly influenced by alumina and silica in the mid-IR wavelengths with stronger correlations in the 2000–500 cm$^{-1}$ (5–20 μm) spectrum. Both metrics also indicate a stronger correlation of ACCUCAST with alumina, which is expected for the major constituent.

### 3.2 Temperature-Dependent Spectroscopic Measurements

Even though the KBr limits the maximum temperature of reflectance measurements, these measurements are useful for obtaining reflectance using the same approach applied for the room-temperature measurements, i.e., as a ratio of intensity measurements of the sample with that obtained for the background KBr, as a function of temperature. Therefore, any non-linearities in detector sensitivity with sample temperature is accounted for both in the sample and background measurements. Moreover, these datasets were used to validate the approach of using relative intensity measurements combined with the room-temperature reflectance spectra to deduce sample reflectance as a function of temperature (Section 2.2, Figure A.4).

Figures 8a–c show the average reflectance measured over three trials for a 5% mass fraction of ACCUCAST, alumina, and silica up to a temperature of 500 °C. At 500 °C, the deviation in reflectance (averaged over all wavelengths) from the mean of the three trials was no larger than 4.55% for any individual trial of the three materials tested. The features and trends observed in the room temperature measurements are present in the temperature-dependent

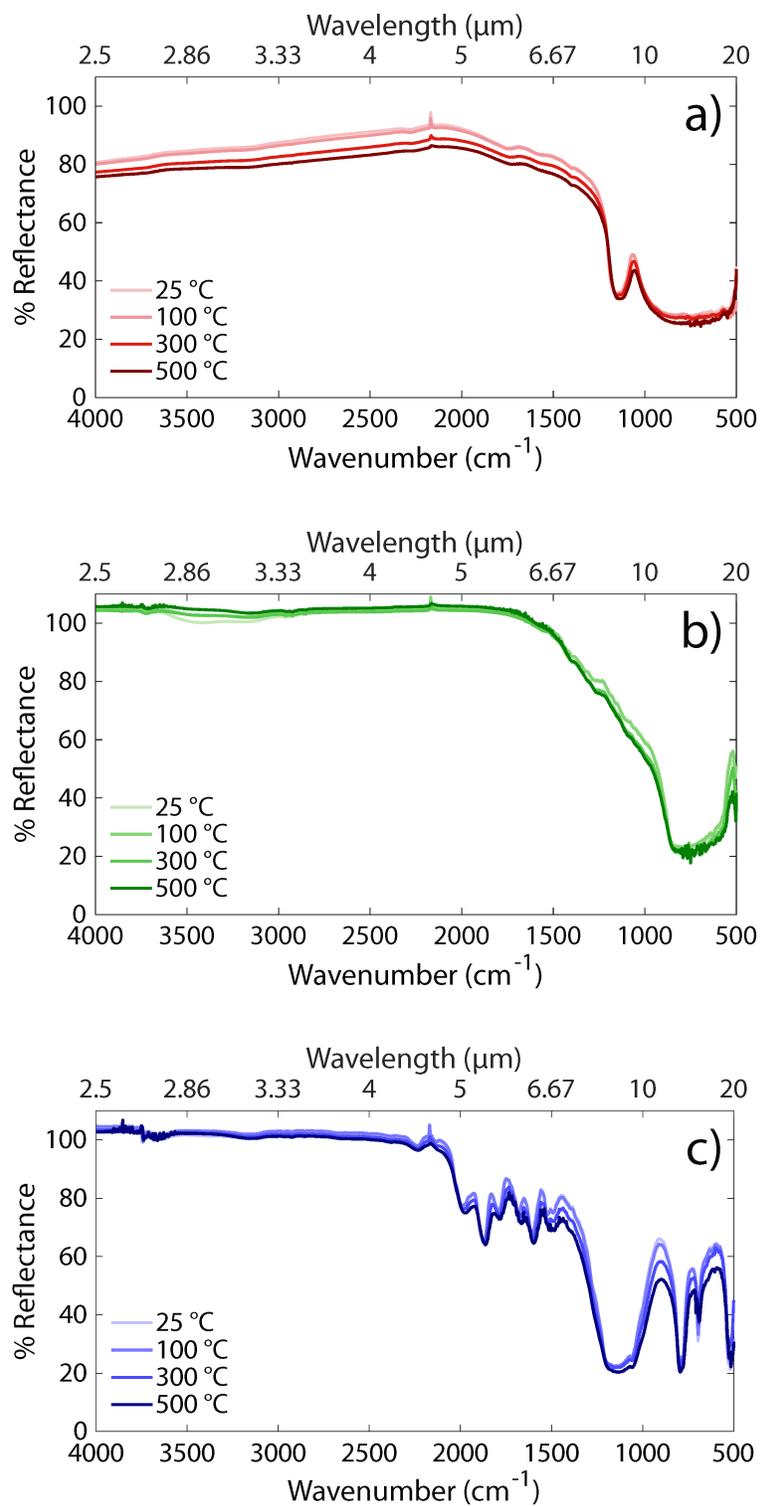

*Figure 8.* Temperature-dependent diffuse reflectance spectra averaged over 3 trials for 5% mass fractions of a) ACCUCAST ID80, b) alumina, and c) silica from 25 °C to 500 °C for a spectral range of 4000–500 cm$^{-1}$ (2.5–20 μm).

measurements, except for the shorter wavelengths (< 4 µm) that do not exhibit the same behavior in alumina and silica. In this range, the absorption band associated with the presence of hydroxyl groups, 3700–2700 cm$^{-1}$ (2.7–3.7 µm), which caused a decrease in room-temperature reflectance of 10–20% (Figure 3), are not seen, indicating these groups are at least partially removed during the heating process. However, the reflectance behavior of ACCUCAST is largely unchanged, which demonstrates that O-H vibrational modes are not responsible for its absorptance in this part of the spectrum.

For ACCUCAST, the reflectance decreases monotonically with increasing temperature from 25 °C to 500 °C (Figure 8a). The peak reflectance at 2150 cm$^{-1}$ (4.65 µm) decreases by 7% at 500 °C compared to the reflectance at 25 °C. For alumina, temperature does not have a very pronounced effect on the reflectance spectrum from 25 °C to 500 °C, except for the moderate increase in reflectance by a maximum of 5% in the spectral range 3700–2700 cm$^{-1}$ (2.7–3.7 µm) attributed to partial loss of surface hydroxyl groups (Figure 8b). For the remainder of the spectrum, the reflectance decreases by less than 4% from 25 °C to 500 °C. In the case of silica (Figure 8c), the reflectance is mostly unaltered by temperature up to 2000 cm$^{-1}$ (5 µm). For larger wavelengths, the effect of temperature becomes more pronounced, with the reflectance decreasing by up to 15% from 25 °C to 500 °C at the local maximum in reflectance at 910 cm$^{-1}$ (11 µm).

Figures 9a–c show reflectance as a function of temperature up to 1000 °C for pure samples (100% mass fraction) of ACCUCAST, alumina, and silica over an extended spectrum of 10000–500 cm$^{-1}$ (1–20 µm), compared to the 4000–500 cm$^{-1}$ (2.5–20 µm) spectrum in Figures 8a-c. For this figure, reflectance was computed by combining relative intensity measurements for each material (Figure A.7) with room-temperature reflectance measurements at 100% mass fraction (Section 2.3). For ACCUCAST at 500 °C, the average deviation between the extrapolated

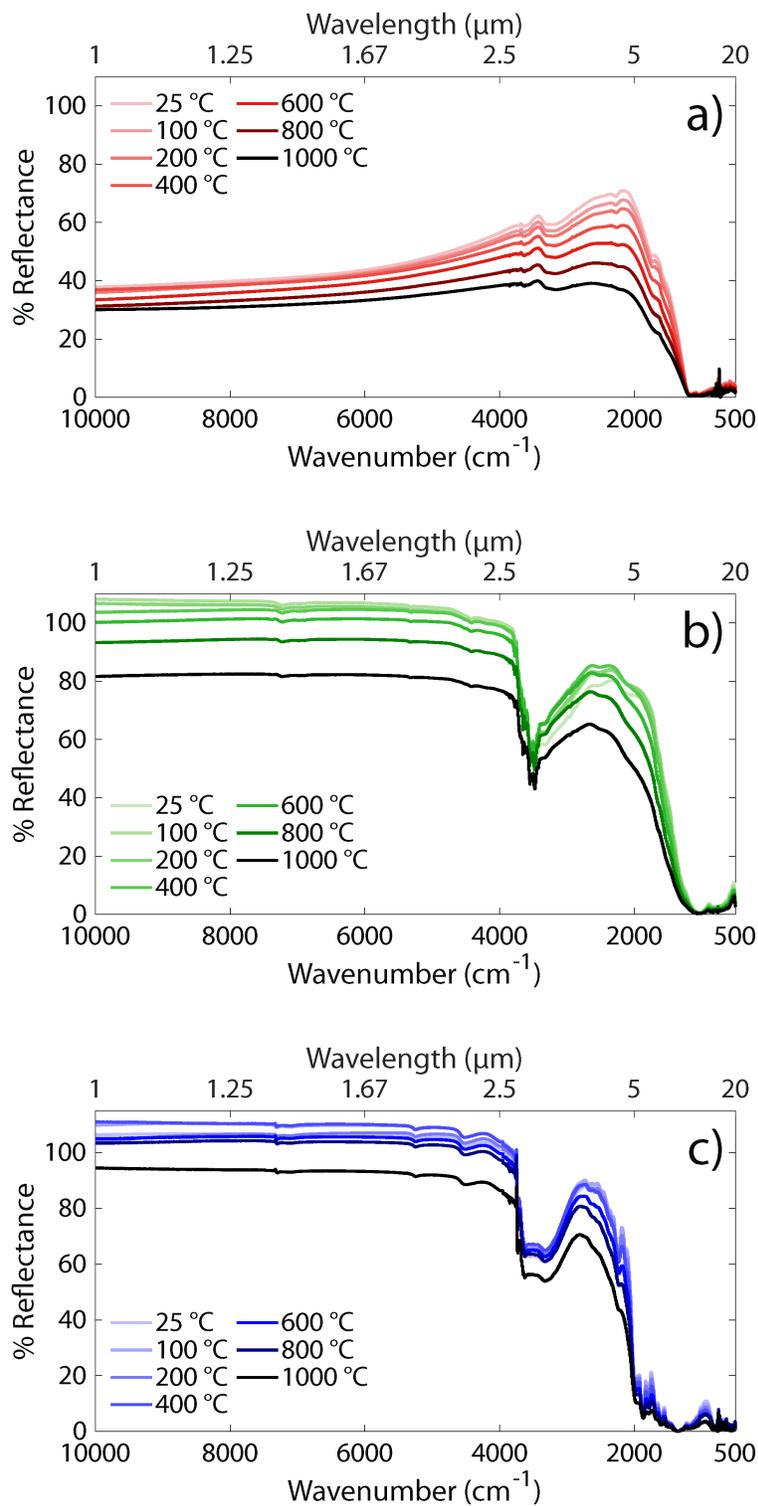

*Figure 9.* Temperature-dependent diffuse reflectance and extrapolated reflectance spectra for 100% mass fractions of a) ACCUCAST ID80, b) alumina, and c) silica from 25 °C to 1000 °C for a spectral range of 10000–500 cm$^{-1}$ (1–20 μm).

reflectance and true reflectance values over the spectrum is 2.26%, and 95% are within 4.44% of the measured reflectance values (Figure A.4). For alumina and silica, the average deviations are similar in magnitude (< 3.3%). Because these results are for the pure samples, the magnitude of reflectance in Figure 9 in the spectrum with high absorption is smaller than that reported in Figure 8. Moreover, the absorption band due to surface hydroxyl groups is more pronounced for the pure samples in Figure 9 as compared to what was shown in Figure 8.

Similar to the trends in Figure 8a, Figure 9a shows that for ACCUCAST, the reflectance decreases, and therefore absorptance increases, monotonically with temperature at all wavelengths. The biggest decrease in reflectance is observed at 2130 cm$^{-1}$ (4.7 μm) where reflectance drops 34.2% from 25 °C to 1000 °C. In the spectral region of highest absorption, 1200–500 cm$^{-1}$ (8.33–20 μm), the effect of temperature is less pronounced because absorption dominates the spectrum. The semi-gray behavior for ACCUCAST in the spectral band 10000–5000 cm$^{-1}$ (1–2 μm) is maintained at all temperatures, and reflectance increases in the spectral band 5000–2000 cm$^{-1}$ (2–5 μm) non-linearly with temperature. Alumina and silica demonstrate a more pronounced decrease in reflectance with temperature from 800 °C to 1000 °C as compared to their reflectance behavior from 25 °C to 600 °C (Figures 9b–c). In the near-IR, 10000–4000 cm$^{-1}$ (1–2.5 μm), reflectance decreases by >23% and >12% for alumina and silica, respectively, going from 25 °C to 1000 °C. Additionally, both materials exhibit gray behavior in the 10000–4000 cm$^{-1}$ (1–2.5 μm) spectrum with high reflectance values (~108% for alumina and ~111% for silica at their peak).

Common infrared spectral features correlated with photon-phonon interactions for solids are broadening and red-shifting (shifting to longer wavelengths) of absorption peaks with increase in temperature [77,78]. While these effects are not very pronounced for any of the materials tested in powder form in this study, the increase in absorption with temperature may be attributed to larger carrier concentrations due to the presence of impurities/defects in the crystal structure of the materials [79]. For single-crystal alumina, prior work has demonstrated an increase in its spectral absorption coefficient as a function of temperature in the near-IR (0.5–6 µm), which matches the measured decrease in reflectance (increase in absorptance) with temperature [80]. More detailed *in-situ* structural and phase characterization needs to be performed together with reflectance measurements to confirm this hypothesis.

To quantify the effects of temperature on radiative properties, Figure 10 shows the thermal emittance (eq. 11) for the three materials from 25 °C to 1000 °C. The pure sample reflectance measurements in Figure 9 were used to obtain a thermal emittance over the spectral range 10000–500 cm$^{-1}$ (1–20 µm). While the absolute values for emittance are specific to the measurement approach and sample morphologies tested in this work, it is illustrative to understand the trends with temperature and compare the emittance of the three materials relative to each other. At 25 °C, the thermal emittance of all three materials overlap at ~0.9. The large thermal emittance at 25 °C is because of the large absorptance/emittance (low reflectance) measured for these materials in the longer wavelengths, where the blackbody emission spectrum is concentrated at 25 °C. The dependency of thermal emittance on temperature is affected by two factors—(1) the increased weightage of spectral emittance in the near-IR with increasing temperature due to the shift in the blackbody spectrum towards shorter wavelengths, and (2) the effect of temperature on the measured reflectance spectra (Figure 9). For all three materials, the decrease in emittance with increase in temperature is attributed to the increased weightage of decreasing absorptance (increasing reflectance) in the shorter wavelengths. Alumina and silica especially exhibit low absorptance (high reflectance) for wavelengths less than 5 µm. Consequently, thermal emittance decreases with temperature by roughly a factor of 2 from 25 °C to 1000 °C. This trend is also consistent with normal total emittance measurements for alumina reported in ref. [81], which show emittance decreasing from 0.8 to 0.4 as temperature increases from 25 °C to 1500 °C. However, for ACCUCAST, thermal emittance initially decreases from 0.91 to 0.64 when temperature increases from 25 °C to 600 °C, and emittance increases thereafter, resulting in a local minimum at 600 °C. This is because ACCUCAST is more absorbing than alumina or silica in the near-IR, and absorptance/emittance increases with temperature (reflectance decreases in Figure 9a) in this spectrum. Combined, this results in an increase in thermal emittance from 600 °C to 1000 °C. A similar effect also occurs in alumina from 800 °C to 1000 °C, while silica demonstrates a decreasing slope at these temperatures.

These results imply that the radiative behavior of ACCUCAST is dictated by alumina and silica at low temperatures, up to ~400 °C. However, it will emit and absorb more radiative energy at higher temperatures, compared to a material composed of purely alumina and silica. Based on the thermal emittance values for ACCUCAST, alumina, and silica, it can also be inferred that to increase the absorptance for temperatures from 25 °C to 400 °C, it may be beneficial to increase the relative amount of silica compared to alumina in the composite mixture. These inferences also mirror the trends observed in the room-temperature reflectance measurements. ACCUCAST is strongly correlated to a mixture of alumina and silica at longer wavelengths, but less so at shorter wavelengths which become more dominant in influencing the thermal emittance at higher temperatures.

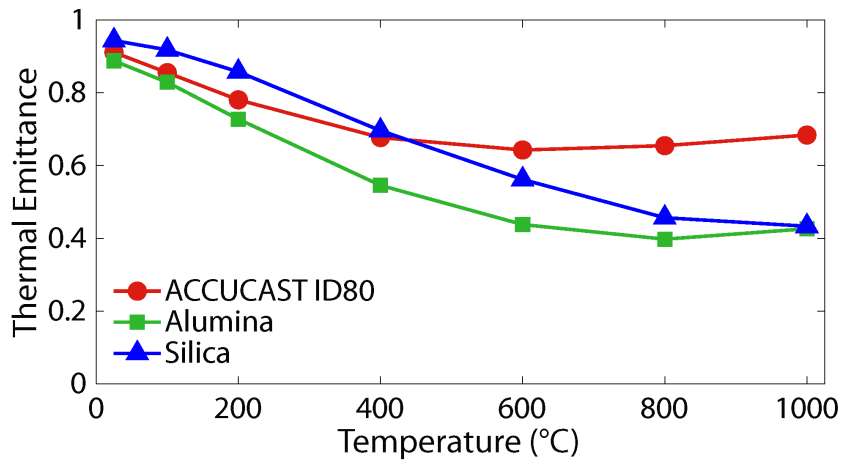

*Figure 10.* Calculated thermal emittance for each material based on reflectance data up to 1000 °C

**3.3 Effects of Thermal Cycling**

Figure 11 presents the effects of thermal cycling on reflectance measured at room temperature, with each material being compared before and after the thermal cycling processes at 500 °C and 1000 °C in air (and vacuum for ACCUCAST). Samples were held at the maximum temperature for 6 hours and reflectance was measured with a mass fraction of 5%. The most significant change in reflectance occurred in the spectrum for ACCUCAST (Figure 11a). Although the change in reflectance after thermal cycling at 500 °C is minimal, a large increase in reflectance (>30%) is observed at 4000 cm$^{-1}$ (2.5 µm) after thermal cycling at 1000 °C, and the reflectance of this sample remains at least 10% above the maximum reflectance of the sample with no treatment up to ~1700 cm$^{-1}$ (5.9 µm). The spectra eventually converge and stay within a few percent <1250 cm$^{-1}$ (>8 µm). Optical microscope images in Figure 12 show a color change in the powder from dark gray to light orange due to thermal cycling in air. Thermal cycling in vacuum produced no discernible color change despite the sample spending the same amount of time at 1000 °C. Additionally, the reflectance of the sample heated in vacuum deviated by no more than 7% from the reflectance of the sample with no treatment. Therefore, the change in radiative properties in ACCUCAST is attributed to oxidation reactions and corresponding phase changes in the constituent oxides at elevated temperatures in the presence of air. Siegel et al. also reported on color changes to sintered bauxite proppants with thermal cycling and attributed this effect to reactions involving iron- and aluminum-titanate species [29]. Moreover, diffuse reflectance measurements made with a UV–Vis shows the trend of increased reflectance (decreased absorptance) continues into the solar spectrum, accounting for the observed color change (Figure A.8).

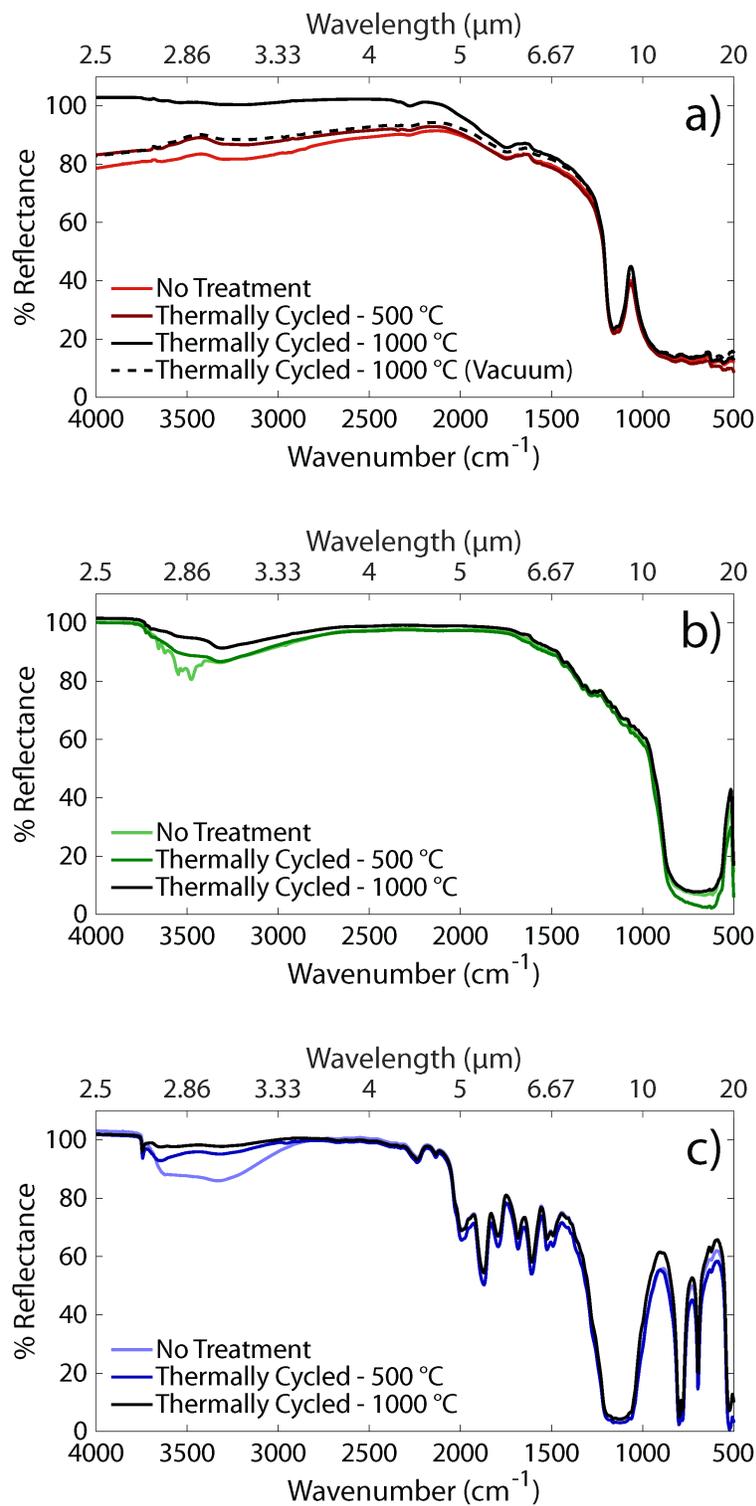

*Figure 11.* Diffuse reflectance spectra before and after thermal cycling at 500 °C and 1000 °C for a) ACCUCAST ID80, b) alumina, and c) silica for a mass fraction of 5% for a spectral range of 4000–500 cm$^{-1}$ (2.5–20 μm).

Therefore, the effect of thermal cycling in air is detrimental to the solar absorptance of directly irradiated CSP receivers with ACCUCAST particles as the heat-transfer fluid. Although it could be speculated that this color change is driven by the oxidation of iron (II,III) oxide—$Fe_3O_4$ (dark gray) + $O_2$ → $Fe_2O_3$ (red-orange) [82,83]—compositional analyses via XRD did not reflect any significant changes before and after thermal cycling (Figure A.1). Because proppant materials are a complex mixture of various metal oxides, more comprehensive, in-situ compositional analysis is essential to isolate the underlying cause and determine mechanisms to control the loss of solar absorptance of these materials with thermal cycling.

Both alumina (Figure 11b) and silica (Figure 11c) display very similar trends before and after thermal cycling. The differences lie in the O-H absorption band, indicating the thermal cycling process removes hydroxyl groups from the samples which does not return even after cooling. The spectra for the samples with no thermal treatment are exceptionally like the thermally cycled spectra $< 3000$ cm$^{-1}$ ($> 3.33$ μm) for alumina and $< 2050$ cm$^{-1}$ ($> 4.9$ μm) for silica. The lack of increased reflectance in the near-IR for alumina and silica also reinforces that another species is driving the change in reflectance in ACCUCAST.

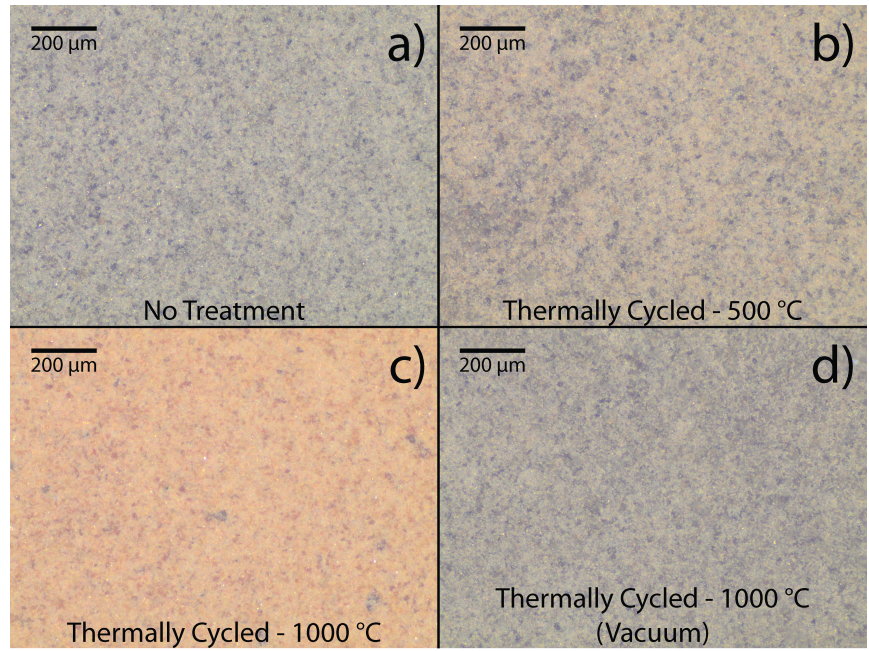

*Figure 12.* Microscope images for ACCUCAST ID80 with a) no thermal cycling, b) thermal cycling at b) 500 °C and c) 1000 °C in air, and d) 1000 °C in vacuum. Samples were held at temperature for 6 hours and imaged at room temperature. Microscope images were acquired at 5X magnification with an exposure time of 300 ms.

## 4. Limitations and Future Work

All measurements reported in this work have been obtained using a DiffusIR diffuse reflectance accessory, which is designed for assessing radiative properties of powdered materials. To transform measured intensities at the detector to a reflectance value, a bed of KBr powder is assumed to be an ideal reflector over the measured spectral range, 10000–500 cm$^{-1}$ (1–20 µm). While KBr is highly transparent at these wavelengths, it is not perfectly transparent, and its transmittance is not completely uniform across all wavelengths [55,67]. Thus, certain reflectance spectra, such as that of alumina, give values above 100% where the sample material is more reflective than KBr. Moreover, reflectance at temperatures beyond the melting point of KBr (730 °C) cannot be obtained as a direct ratio between the spectral intensities measured for the sample and KBr at any temperature. Reflectance values at temperatures above 500 °C are instead estimated by using ratios of reflected intensity for the pure samples at different temperatures. Both the direct approach and the extrapolation of relative intensity to a reflectance using the room-temperature measurements match reasonably well up to 500 °C (Figure A.4). However, up to 1000 °C, the deviation is expected to be larger. Any non-linearities in the detector response due to increase in thermal emission from the sample at higher temperatures can be better accounted for with a diffuse reflectance standard that can withstand the highest temperatures. Such a standard is not yet commercially available or reported in the literature and will be a useful future research direction. In this study, reflectance and thermal emittance are reported for all the materials tested; however, these properties are more representative of an ensemble of particles as compared to intrinsic material properties. Future work will consider the development of inverse modeling tools to extract materials-specific, temperature-dependent radiative properties of the materials [54]. Thermal cycling measurements for ACCUCAST demonstrate the influences of radiative property measurements in air as compared to a vacuum environment. While temperature-dependent radiative property data for all materials is reported under evacuated conditions, measurements in air will be useful to consider. Finally, to enable comparison of measurements across independent investigations, a more comprehensive approach is required towards a full-fledged error propagation for all measured quantities, beyond the random error quantified in this work, and will be considered in future work [84].

## 5. Summary and Conclusions

Diffuse reflectance was measured over the spectral range 10000–500 cm$^{-1}$ (1–20 µm) for three ceramic powders—ACCUCAST ID80, alumina, and silica—from 25 °C to 1000 °C using an FTIR, a diffuse reflectance accessory, and a heated stage. DRS measurements performed at room temperature showed that higher concentrations of sample material result in more absorption in the particulate medium composed of the sample and KBr. All three materials were most reflective up to around 2000 cm$^{-1}$ (5 µm), with strong absorption features in the mid-IR that correspond to different lattice vibrational modes. The absorption features of ACCUCAST were correlated to absorption features observed in alumina and silica, its two major constituents, and with a binary mixture of alumina and silica. In the mid-IR wavelengths (5–20 µm), the reflectance spectrum of ACCUCAST is most strongly correlated with alumina and the binary mixture compared to silica. However, in the near-IR wavelengths (1–5 µm), the absorption behavior of ACCUCAST is distinct from those obtained for alumina, silica, and the binary mixture. This indicates contributions of radiative behavior from another chemical species, possibly iron oxide, present in the composite mixture.

Heated-stage DRS measurements showed the dependence of each material's radiative properties as a function of temperature. The effect of temperature was found to depend on the material and wavelength, but generally a decrease in reflectance was observed with temperature. The near-IR wavelengths were weakly dependent on temperature up to 600 °C, and the largest changes were observed in the mid-IR where most absorption features are present. To determine reflectance above the operating temperature (melting point) of the background material KBr, a new technique was applied that calculated the relative change in intensity for the same material at different temperatures and applied this as a scaling factor to the measured reflectance of the same material at room-temperature. This technique proved to be effective as it resulted in small errors (<5%) for temperatures up to 500 °C, and is a viable approach to compute reflectance in the absence of a diffuse background material that is stable up to 1000 °C. With this technique thermal emittance was calculated using the thermal emission spectrum from 1–20 μm (eq. 11) and found to decrease with temperature up to 600 °C for all materials. However, the competing effects of the blackbody emission spectrum shifting towards shorter wavelengths and spectral reflectance decreasing with temperature led to a more complex behavior. Ultimately, thermal emittance increased starting at 600 °C and 800 °C ACCUCAST and alumina, respectively. Thermal cycling of ACCUCAST showed that reflectance increases when it is exposed to high temperatures for prolonged periods of time in an air environment. The biggest changes came in the near-IR as the thermal cycling temperature was increased from 500 °C to 1000 °C, which also coincided with a color change in the material from gray to light orange. Contrastingly, no color change or significant increase in reflectance was observed when the ACCUCAST was heated in vacuum up to 1000 °C, indicating oxidation likely drives the change in radiative behavior of ACCUCAST in the visible spectrum.

Collectively, this study reports on crucial advancements to experimental methods development, reflectance measurements, and interpretations of temperature-dependent radiative properties. In addition to quantifying spectral- and temperature-dependent properties, interpreted results also provide insight on future material compositions to improve thermal absorptance/emittance behavior. The net effect of absorption in the visible spectrum and re-radiation in the infrared spectrum will ultimately determine the effectiveness of ACCUCAST as a heat-transfer medium for CSP applications. Thus, the reported spectral radiative property data in the IR wavelengths and its temperature dependencies will be crucial towards optimizing the performance of these systems.

## Acknowledgements

The authors acknowledge the financial support provided by startup funds from the Department of Mechanical Engineering and the College of Engineering, and partial funding for Mayer from the Carbon Neutrality Acceleration Program through the Graham Sustainability Institute at the University of Michigan. The authors also acknowledge the UM Battery Lab for providing access to equipment for materials characterization along with technical assistance. Graduate student Sanjay Bharati is acknowledged for assistance with sample preparation and running additional reflectance tests to help quantify repeatability.

**Appendix**

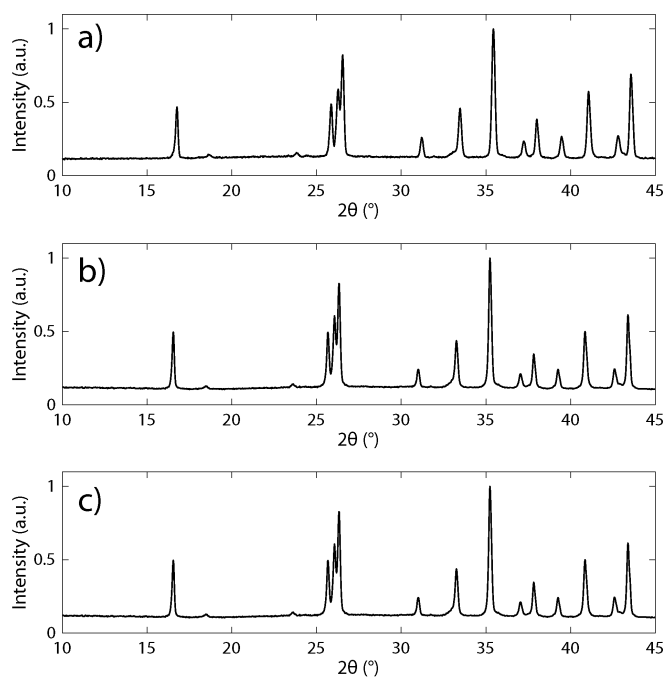

*Figure A.1.* PXRD measurements of ACCUCAST ID80 a) as received, b) after the grinding and sieving process, and c) after thermal cycling at 1000 °C in air. Relative peak intensities suggest the material composition is unchanged after the particle size reduction process and thermal cycling.

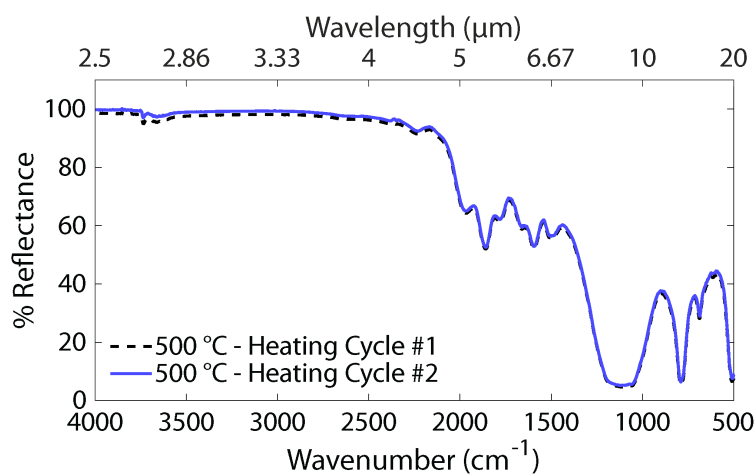

*Figure A.2.* Diffuse reflectance spectra for silica at 500 °C during the first and second heating cycles of the sample for a mass fraction of 5% and a spectral range of 4000–500 cm$^{-1}$ (2.5–20 µm). The reflectance values of the spectra are within 3% of each other at all measured wavelengths.

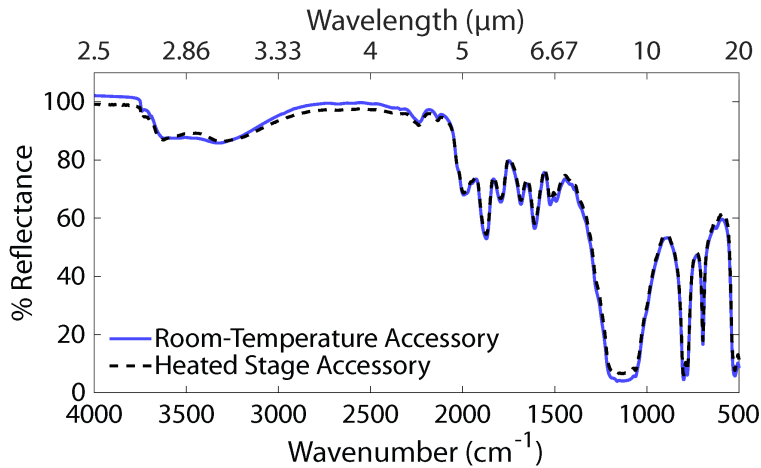

*Figure A.3.* Diffuse reflectance spectra for silica in the room-temperature and heated stage accessories for a mass fraction of 5% and a spectral range of 4000–500 cm$^{-1}$ (2.5–20 µm). The reflectance values of the spectra are within 3.5% of each other at all measured wavelengths.

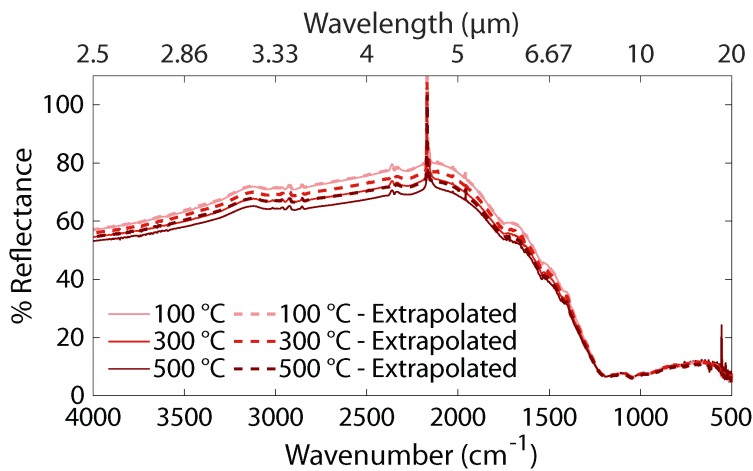

*Figure A.4.* Temperature-dependent diffuse reflectance spectra for ACCUCAST for a mass fraction of 100% over a spectral range of 4000–500 cm$^{-1}$ (2.5–20 µm). Measured reflectance values are shown as solid lines, and extrapolated reflectance values at the same temperatures are shown as dashed lines. The extrapolated reflectance values deviate from the measured values more as temperature increases. At 500 °C, the average deviation over the spectrum is 2.26% and 95% of extrapolated values are within 4.44% of the measured reflectance values. Note: the spike in reflectance at 2170 cm-1 (4.6 µm) is an artifact of a KBr absorption feature that appears after heating at 500 °C. The exact origins of this peak are unknown.

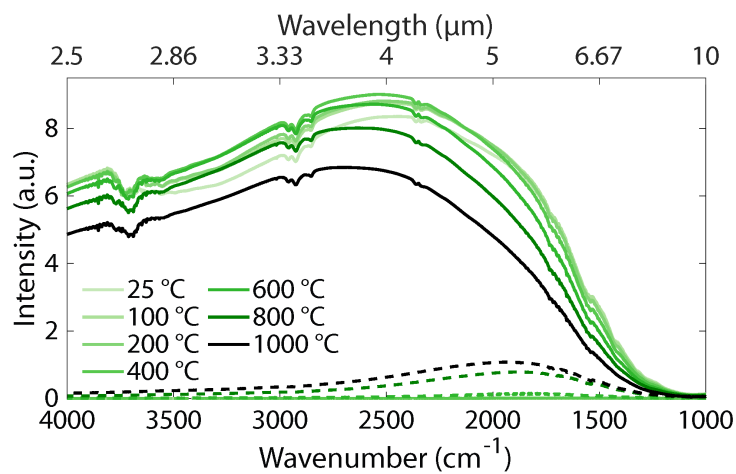

*Figure A.5*. Temperature-dependent reflected intensity spectra (solid lines) and emission spectra (dashed lines) for pure alumina over a spectral range of 4000–500 cm$^{-1}$ (2.5–20 µm). The emission spectra are acquired by turning off the IR source beam and measuring the signal. Even at 1000 °C, the emission spectrum is only a small fraction of the total intensity.

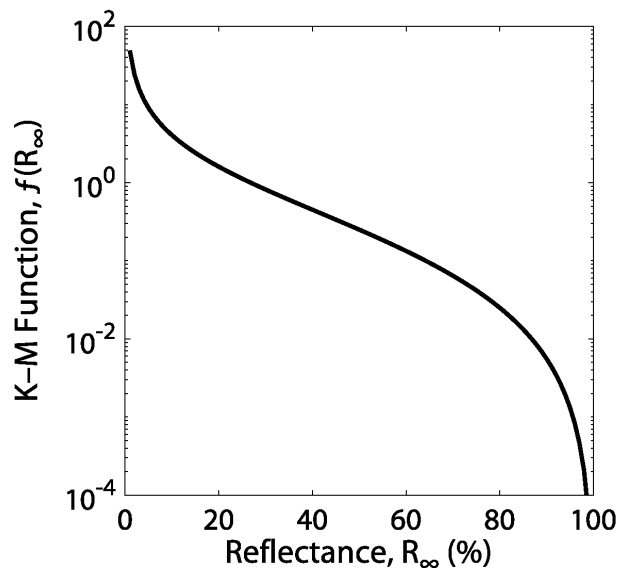

*Figure A.6*. The Kubelka–Munk function plotted on a log scale with respect to reflectance. The values of the K–M function rapidly increase as reflectance approaches 0%. For high values of reflectance, the K–M function tends toward a value of zero.

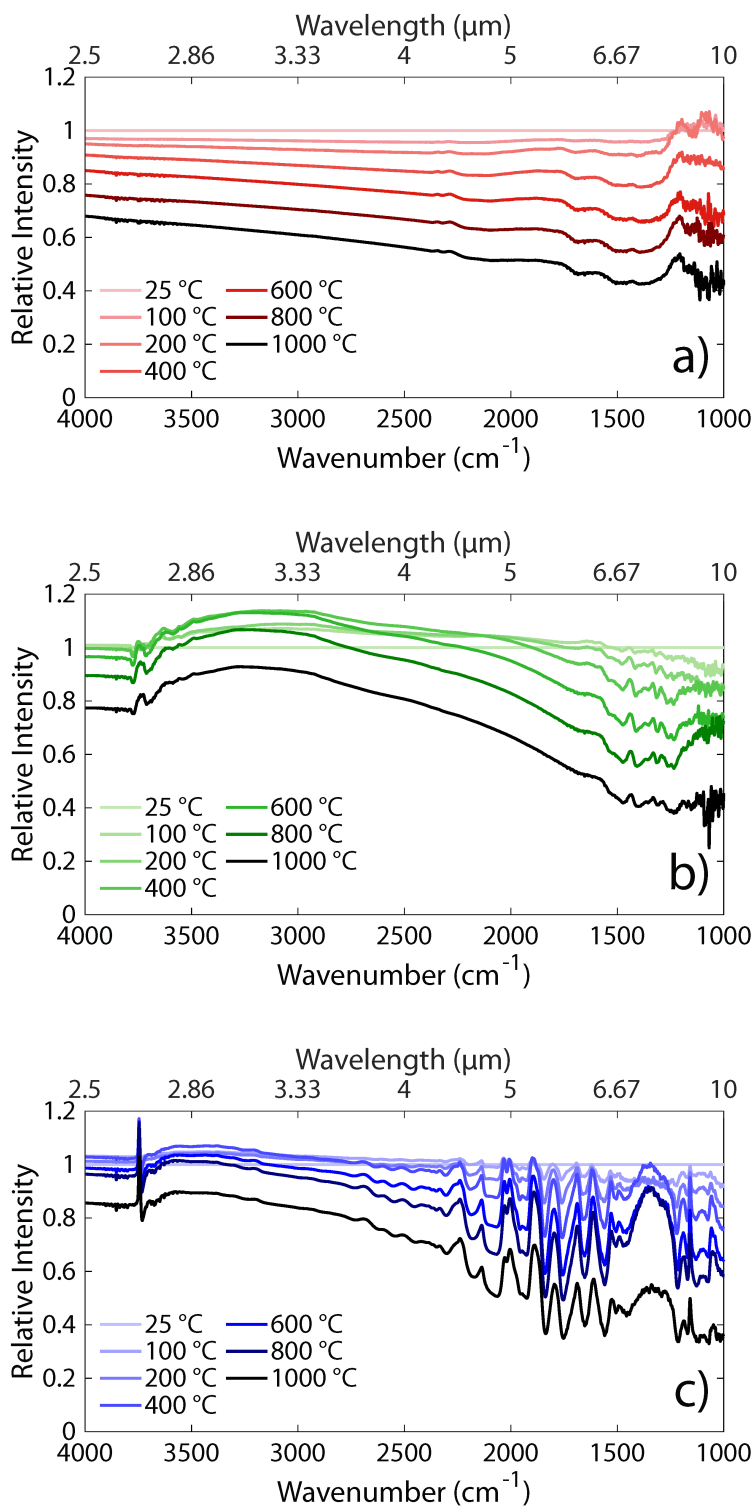

*Figure A.7.* Temperature-dependent relative intensity spectra for a) ACCUCAST ID80, b) alumina, and c) silica for a mass fraction of 100% from 25 °C to 1000 °C for a spectral range of 4000–1000 cm$^{-1}$ (2.5–10 μm)

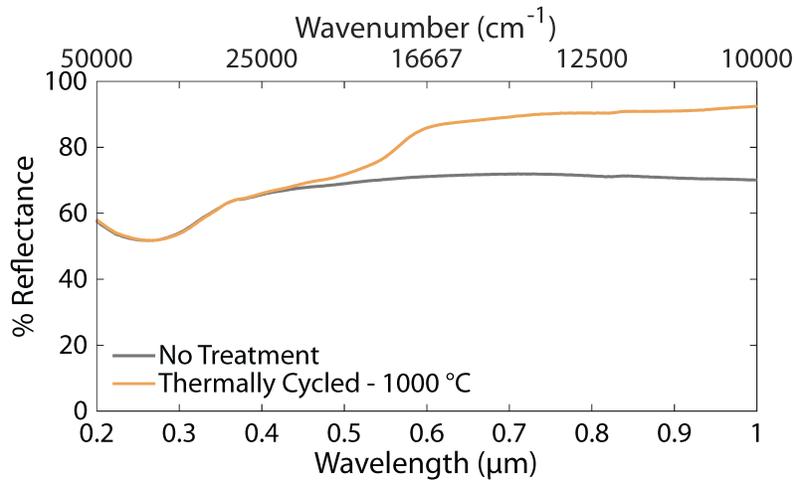
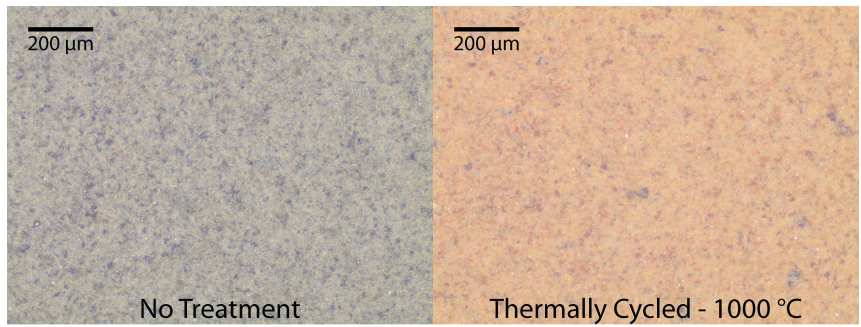

*Figure A.8*. UV–Vis diffuse reflectance spectra and microscope images for ACCUCAST ID80 with no thermal cycling and thermal cycling at 1000 °C in air. Spectra were measured for a mass fraction of 10% and a spectral range of 50000–10000 cm$^{-1}$ (0.2–1 µm). Microscope images were acquired at 5X magnification with an exposure time of 300 ms. The thermally cycled sample exhibits a greater reflectance for wavelengths >0.5 µm, resulting in a color change. The sample with no heat treatment exhibits largely gray behavior across the visible spectrum. Reflectance spectra are the average of three trials.

# References

[1]  T. Tan, Y. Chen, Review of study on solid particle solar receivers, Renew. Sustain. Energy Rev. 14 (2010) 265–276. https://doi.org/10.1016/j.rser.2009.05.012.

[2]  C.K. Ho, A review of high-temperature particle receivers for concentrating solar power, Appl. Therm. Eng. 109 (2016) 958–969. https://doi.org/10.1016/j.applthermaleng.2016.04.103.

[3]  M. Ebert, L. Amsbeck, J. Rheinländer, B. Schlögl-Knothe, S. Schmitz, M. Sibum, R. Uhlig, R. Buck, Operational experience of a centrifugal particle receiver prototype, AIP Conf. Proc. 2126 (2019) 1–8. https://doi.org/10.1063/1.5117530.

[4]  E. Alonso, M. Romero, Review of experimental investigation on directly irradiated particles solar reactors, Renew. Sustain. Energy Rev. 41 (2015) 53–67. https://doi.org/10.1016/j.rser.2014.08.027.

[5]  T. Kodama, S. Bellan, N. Gokon, H.S. Cho, Particle reactors for solar thermochemical processes, Sol. Energy. (2017). https://doi.org/10.1016/j.solener.2017.05.084.

[6]  D.M. Fabian, S. Hu, N. Singh, F.A. Houle, T. Hisatomi, K. Domen, F.E. Osterloh, S. Ardo, Particle suspension reactors and materials for solar-driven water splitting, Energy Environ. Sci. 8 (2015) 2825–2850. https://doi.org/10.1039/C5EE01434D.

[7]  S. Chen, T. Takata, K. Domen, Particulate photocatalysts for overall water splitting, Nat. Rev. Mater. 2 (2017) 17050. https://doi.org/10.1038/natrevmats.2017.50.

[8]  M. Qian, Metal Powder for Additive Manufacturing, JOM. 67 (2015) 536–537. https://doi.org/10.1007/s11837-015-1321-z.

[9]  V. Bhavar, P. Kattire, V. Patil, S. Khot, K. Gujar, R. Singh, A review on powder bed fusion technology of metal additive manufacturing, in: Addit. Manuf. Handb. Prod. Dev. Def. Ind., 2017: pp. 251–261. https://doi.org/10.1201/9781315119106.

[10] M.F. Modest, Radiative exchange between gray, diffuse surfaces, in: Radiat. Heat Transf., Second, 2003: pp. 160–196.

[11] M.F. Modest, The Radiative Transfer Equation in Participating Media(RTE), in: Radiat. Heat Transf., Second, Academic Press, 2003: pp. 279–302.

[12] X.Q. Cao, R. Vassen, D. Stoever, Ceramic materials for thermal barrier coatings, J. Eur. Ceram. Soc. 24 (2004) 1–10. https://doi.org/10.1016/S0955-2219(03)00129-8.

[13] S. Zhao, X. Li, X. Huai, K. Cheng, X. Zhou, Experimental study on the scattering and absorption coefficients of thermal barrier coatings at elevated temperatures, Int. J. Heat Mass Transf. 121 (2018) 900–910. https://doi.org/10.1016/j.ijheatmasstransfer.2018.01.028.

[14] B.J. Hathaway, R. Bala Chandran, A.C. Gladen, T.R. Chase, J.H. Davidson, Demonstration of a Solar Reactor for Carbon Dioxide Splitting via the Isothermal Ceria Redox Cycle and Practical Implications, Energy & Fuels. 30 (2016) 6654–6661. https://doi.org/10.1021/acs.energyfuels.6b01265.

[15] P. Furler, J. Scheffe, D. Marxer, M. Gorbar, A. Bonk, U. Vogt, A. Steinfeld, Thermochemical $CO_2$ splitting via redox cycling of ceria reticulated foam structures with dual-scale porosities, Phys. Chem. Chem. Phys. 16 (2014) 10503. https://doi.org/10.1039/c4cp01172d.

[16] C.L. Muhich, B.D. Ehrhart, I. Al-Shankiti, B.J. Ward, C.B. Musgrave, A.W. Weimer, A review and perspective of efficient hydrogen generation via solar thermal water splitting, Wiley Interdiscip. Rev. Energy Environ. 5 (2016) 261–287.

https://doi.org/10.1002/wene.174.
[17] B.-M. Zhang, S.-Y. Zhao, X.-D. He, Experimental and theoretical studies on high-temperature thermal properties of fibrous insulation, J. Quant. Spectrosc. Radiat. Transf. 109 (2008) 1309–1324. https://doi.org/10.1016/j.jqsrt.2007.10.008.
[18] A. Banerjee, R. Bala Chandran, J.H. Davidson, Experimental investigation of a reticulated porous alumina heat exchanger for high temperature gas heat recovery, Appl. Therm. Eng. 75 (2015) 889–895. https://doi.org/10.1016/j.applthermaleng.2014.10.033.
[19] R. Bala Chandran, R.M. De Smith, J.H. Davidson, Model of an integrated solar thermochemical reactor/reticulated ceramic foam heat exchanger for gas-phase heat recovery, Int. J. Heat Mass Transf. 81 (2015) 404–414. https://doi.org/10.1016/j.ijheatmasstransfer.2014.10.053.
[20] T. Fend, W. Völker, R. Miebach, O. Smirnova, D. Gonsior, D. Schöllgen, P. Rietbrock, Experimental investigation of compact silicon carbide heat exchangers for high temperatures, Int. J. Heat Mass Transf. 54 (2011) 4175–4181. https://doi.org/10.1016/j.ijheatmasstransfer.2011.05.028.
[21] J. Martinek, Z. Ma, Granular Flow and Heat Transfer Study in a Near-Blackbody Enclosed Particle Receiver, in: Proc. ASME 2014 8th Int. Conf. Energy Sustain., 2014. https://doi.org/10.1115/ES2014-6393.
[22] Z. Ma, M. Mehos, G. Glatzmaier, B.B. Sakadjian, Development of a Concentrating Solar Power System Using Fluidized-bed Technology for Thermal Energy Conversion and Solid Particles for Thermal Energy Storage, Energy Procedia. 69 (2015) 1349–1359. https://doi.org/10.1016/j.egypro.2015.03.136.
[23] K.J. Albrecht, H.F. Laubscher, M.D. Carlson, C.K. Ho, Development and Testing of a 20 kW Moving Packed-Bed Particle-To-sCO2 Heat Exchanger and Test Facility, in: ASME 2021 15th Int. Conf. Energy Sustain., American Society of Mechanical Engineers, 2021: pp. 1–6. https://doi.org/10.1115/ES2021-64050.
[24] C.K. Ho, J. Christian, J. Yellowhair, S. Jeter, M. Golob, C. Nguyen, K. Repole, S. Abdel-Khalik, N. Siegel, H. Al-Ansary, A. El-Leathy, B. Gobereit, Highlights of the high-temperature falling particle receiver project: 2012 - 2016, in: 2017: p. 030027. https://doi.org/10.1063/1.4984370.
[25] F. Crespi, G. Gavagnin, D. Sánchez, G.S. Martínez, Supercritical carbon dioxide cycles for power generation: A review, Appl. Energy. 195 (2017) 152–183. https://doi.org/10.1016/j.apenergy.2017.02.048.
[26] A.L. Ávila-Marín, Volumetric receivers in Solar Thermal Power Plants with Central Receiver System technology: A review, Sol. Energy. 85 (2011) 891–910. https://doi.org/10.1016/J.SOLENER.2011.02.002.
[27] G. Zanganeh, A. Pedretti, S. Zavattoni, M. Barbato, A. Steinfeld, Packed-bed thermal storage for concentrated solar power – Pilot-scale demonstration and industrial-scale design, Sol. Energy. 86 (2012) 3084–3098. https://doi.org/10.1016/J.SOLENER.2012.07.019.
[28] C.K. Ho, A.R. Mahoney, A. Ambrosini, M. Bencomo, A. Hall, T.N. Lambert, Characterization of Pyromark 2500 Paint for High-Temperature Solar Receivers, J. Sol. Energy Eng. 136 (2014) 1–4. https://doi.org/10.1115/1.4024031.
[29] N.P. Siegel, M.D. Gross, R. Coury, The Development of Direct Absorption and Storage Media for Falling Particle Solar Central Receivers, J. Sol. Energy Eng. 137 (2015) 041003. https://doi.org/10.1115/1.4030069.

[30] M. Mehos, C. Turchi, J. Vidal, M. Wagner, Z. Ma, C. Ho, W. Kolb, C. Andraka, A. Kruizenga, Concentrating Solar Power Gen3 Demonstration Roadmap, Golden, CO (United States), Alburquerque, NM, Livermore, CA, 2017.

[31] Y.S. Touloukian, D.P. DeWitt, Thermal Radiative Properties - Metallic Elements and Alloys, in: Thermophys. Prop. Matter - TPRC Data Ser. (Volume 7), 1970.

[32] Y.S. Touloukian, D.P. DeWitt, Thermal Radiative Properties - Nonmetallic Solids, in: Thermophys. Prop. Matter - TPRC Data Ser. (Volume 8), 1972.

[33] Y.S. Touloukian, D.P. DeWitt, Thermal Radiative Properties - Coatings, in: Thermophys. Prop. Matter - TPRC Data Ser. (Volume 9), 1972.

[34] A.M. Baldridge, S.J. Hook, C.I. Grove, G. Rivera, The ASTER spectral library version 2.0, Remote Sens. Environ. 113 (2009) 711–715. https://doi.org/10.1016/j.rse.2008.11.007.

[35] S.K. Meerdink, S.J. Hook, D.A. Roberts, E.A. Abbott, The ECOSTRESS spectral library version 1.0, Remote Sens. Environ. 230 (2019) 111196. https://doi.org/10.1016/j.rse.2019.05.015.

[36] P.R. Christensen, J.L. Bandfield, V.E. Hamilton, D.A. Howard, M.D. Lane, J.L. Piatek, S.W. Ruff, W.L. Stefanov, A thermal emission spectral library of rock-forming minerals, J. Geophys. Res. Planets. 105 (2000) 9735–9739. https://doi.org/10.1029/1998JE000624.

[37] W.E. Vargas, G.A. Niklasson, Optical properties of nano-structured dye-sensitized solar cells, Sol. Energy Mater. Sol. Cells. 69 (2001) 147–163. https://doi.org/10.1016/S0927-0248(00)00388-3.

[38] S. Hore, C. Vetter, R. Kern, H. Smit, A. Hinsch, Influence of scattering layers on efficiency of dye-sensitized solar cells, Sol. Energy Mater. Sol. Cells. 90 (2006) 1176–1188. https://doi.org/10.1016/j.solmat.2005.07.002.

[39] J. Kuhn, T. Gleissner, M.C. Arduini-Schuster, S. Korder, J. Fricke, Integration of mineral powders into SiO2 aerogels, J. Non. Cryst. Solids. 186 (1995) 291–295. https://doi.org/10.1016/0022-3093(95)00067-4.

[40] J.-J. Zhao, Y.-Y. Duan, X.-D. Wang, X.-R. Zhang, Y.-H. Han, Y.-B. Gao, Z.-H. Lv, H.-T. Yu, B.-X. Wang, Optical and radiative properties of infrared opacifier particles loaded in silica aerogels for high temperature thermal insulation, Int. J. Therm. Sci. 70 (2013) 54–64. https://doi.org/10.1016/j.ijthermalsci.2013.03.020.

[41] X.-H. Gao, C.-B. Wang, Z.-M. Guo, Q.-F. Geng, W. Theiss, G. Liu, Structure, optical properties and thermal stability of Al2O3-WC nanocomposite ceramic spectrally selective solar absorbers, Opt. Mater. (Amst). 58 (2016) 219–225. https://doi.org/10.1016/j.optmat.2016.05.037.

[42] M.S. Prasad, B. Mallikarjun, M. Ramakrishna, J. Joarder, B. Sobha, S. Sakthivel, Zirconia nanoparticles embedded spinel selective absorber coating for high performance in open atmospheric condition, Sol. Energy Mater. Sol. Cells. 174 (2018) 423–432. https://doi.org/10.1016/j.solmat.2017.09.032.

[43] J.W. Griffin, K.A. Stahl, R.B. Pettit, Optical properties of solid particle receiver materials. I. Angular scattering and extinction characteristics of Norton Masterbeads®, Sol. Energy Mater. 14 (1986) 395–416. https://doi.org/10.1016/0165-1633(86)90062-6.

[44] K.A. Stahl, J.W. Griffin, R.B. Pettit, Optical properties of solid particle receiver materials. II. Diffuse reflectance of Norton Masterbeads® at elevated temperatures, Sol. Energy Mater. 14 (1986) 417–425. https://doi.org/10.1016/0165-1633(86)90063-8.

[45] C. Ho, J. Christian, D. Gill, A. Moya, S. Jeter, S. Abdel-Khalik, D. Sadowski, N. Siegel,

H. Al-Ansary, L. Amsbeck, B. Gobereit, R. Buck, Technology advancements for next generation falling particle receivers, in: Energy Procedia, Elsevier Ltd, 2014: pp. 398–407. https://doi.org/10.1016/j.egypro.2014.03.043.

[46] C. Chen, C. Yang, D. Ranjan, P.G. Loutzenhiser, Z.M. Zhang, Spectral Radiative Properties of Ceramic Particles for Concentrated Solar Thermal Energy Storage Applications, Int. J. Thermophys. 41 (2020) 152. https://doi.org/10.1007/s10765-020-02733-5.

[47] F. Nie, Z. Cui, F. Bai, Z. Wang, Properties of solid particles as heat transfer fluid in a gravity driven moving bed solar receiver, Sol. Energy Mater. Sol. Cells. 200 (2019) 110007. https://doi.org/10.1016/j.solmat.2019.110007.

[48] CARBO, CARBOBEAD - High-performance ceramic media, (2019).

[49] CARBO, High-performance ceramic casting media, (2020). https://carboceramics.com/getmedia/d950de6b-c6a2-45bf-af4e-e2409a889cbc/CARBO-ACCUCAST-Technical-Data-Sheet-1001_340.pdf?ext=.pdf (accessed July 21, 2021).

[50] O. Faix, J.H. Böttcher, The influence of particle size and concentration in transmission and diffuse reflectance spectroscopy of wood, Holz Als Roh- Und Werkst. 50 (1992) 221–226. https://doi.org/10.1007/BF02650312.

[51] A.B. Ertel, C.S. Brauer, R.L. Richardson, R.G. Tonkyn, T.L. Myers, T.A. Blake, T.J. Johnson, Y.-F. Su, Quantitative reflectance spectra of solid powders as a function of particle size, Appl. Opt. Vol. 54, Issue 15, Pp. 4863-4875. 54 (2015) 4863–4875. https://doi.org/10.1364/AO.54.004863.

[52] R.W. Frei, H. Zeitlin, Diffuse Reflectance Spectroscopy, C R C Crit. Rev. Anal. Chem. 2 (1971) 179–246. https://doi.org/10.1080/10408347108542764.

[53] T. Eickhoff, P. Grosse, W. Theiss, Diffuse reflectance spectroscopy of powders, Anal. Chim. Acta. 240 (1990) 229–233. https://doi.org/10.1016/s0003-2670(00)86541-1.

[54] S. Bock, C. Kijatkin, D. Berben, M. Imlau, Absorption and Remission Characterization of Pure, Dielectric (Nano-)Powders Using Diffuse Reflectance Spectroscopy: An End-To-End Instruction, Appl. Sci. 9 (2019) 4933. https://doi.org/10.3390/app9224933.

[55] ThorLabs, Optical Substrates - Potassium Bromide, (n.d.).

[56] Thermo Scientific, Introduction To FT-IR Sample Handling, (n.d.).

[57] PIKE Technologies Inc., DiffusIR – Research Grade Diffuse Reflectance Accessory, (2020).

[58] Royal Society of Chemistry, The Merck Index Online - Potassium Bromide, (2013).

[59] B.H. Zimm, J.E. Mayer, Vapor Pressures, Heats of Vaporization, and Entropies of Some Alkali Halides, J. Chem. Phys. 12 (1944) 362–369. https://doi.org/10.1063/1.1723958.

[60] Agilent Technologies, DS40M Pump - User Manual, (2016).

[61] Knight Optical, Safety Data Sheet - Zinc Selenide Optical Crystal, (2020).

[62] A. Springsteen, Standards for the measurement of diffuse reflectance – an overview of available materials and measurement laboratories, Anal. Chim. Acta. 380 (1999) 379–390. https://doi.org/10.1016/S0003-2670(98)00576-5.

[63] K.C. Weston, A.C. Reynolds, D.W. Drago, Radiative Transfer in Highly Scattering Materials – Numerical Solution and Evaluation of Approximate Analytical Solutions, 1974.

[64] H.G. Hecht, The Interpretation of Diffuse Reflectance Spectra, J. Res. Natl. Bur. Stand. Sect. A Phys. Chem. 80A (1976) 567. https://doi.org/10.6028/jres.080A.056.

[65] A.A. Christy, O.M. Kvalheim, R.A. Velapoldi, Quantitative analysis in diffuse reflectance

spectrometry: A modified Kubelka–Munk equation, Vib. Spectrosc. 9 (1995) 19–27. https://doi.org/10.1016/0924-2031(94)00065-O.

[66] M. Milosevic, S.L. Berets, A Review of FT-IR Diffuse Reflection Sampling Considerations, Appl. Spectrosc. Rev. 37 (2002) 347–364. https://doi.org/10.1081/ASR-120016081.

[67] H.H. Li, Refractive index of alkaline earth halides and its wavelength and temperature derivatives, J. Phys. Chem. Ref. Data. 9 (1980) 161–290. https://doi.org/10.1063/1.555616.

[68] M.F. Modest, Radiative Properties of Real Surfaces, in: Radiat. Heat Transf., Third, 2003: pp. 61–128.

[69] J.. Anderson, K.. Wickersheim, Near infrared characterization of water and hydroxyl groups on silica surfaces, Surf. Sci. 2 (1964) 252–260. https://doi.org/10.1016/0039-6028(64)90064-0.

[70] M.K. Collins, The Use of Diffuse Reflectance Infrared Spectroscopy in the Study of Alumina, Kansas State University, 1986.

[71] E.A. Schatz, Effect of Pressure on the Reflectance of Compacted Powders, J. Opt. Soc. Am. 56 (1966) 389. https://doi.org/10.1364/JOSA.56.000389.

[72] R. Hanna, Infrared Absorption Spectrum of Silicon Dioxide, J. Am. Ceram. Soc. 48 (1965) 595–599. https://doi.org/10.1111/j.1151-2916.1965.tb14680.x.

[73] A.M. Efimov, Quantitative IR spectroscopy: Applications to studying glass structure and properties, J. Non. Cryst. Solids. 203 (1996) 1–11. https://doi.org/10.1016/0022-3093(96)00327-4.

[74] S.W. Hwang, A. Umar, G.N. Dar, S.H. Kim, R.I. Badran, Synthesis and Characterization of Iron Oxide Nanoparticles for Phenyl Hydrazine Sensor Applications, Sens. Lett. 12 (2014) 97–101. https://doi.org/10.1166/sl.2014.3224.

[75] D. Walter, G. Buxbaum, W. Laqua, The mechanism of the thermal transformation from goethite to hematite, J. Therm. Anal. Calorim. 63 (2001) 733–748. https://doi.org/10.1023/A:1010187921227.

[76] Y. Dodge, Spearman Rank Correlation Coefficient, in: Concise Encycl. Stat., Springer New York, New York, NY, 2008: pp. 502–505. https://doi.org/10.1007/978-0-387-32833-1_379.

[77] R.G. Burns, Mineralogical Applications of Crystal Field Theory, (1993). https://doi.org/10.1017/CBO9780511524899.

[78] V.K. Bityukov, V.A. Petrov, Absorption Coefficient of Molten Aluminum Oxide in Semitransparent Spectral Range, Appl. Phys. Res. 5 (2013) p51. https://doi.org/10.5539/APR.V5N1P51.

[79] R.B. Singer, T.L. Roush, Effects of temperature on remotely sensed mineral absorption features, J. Geophys. Res. Solid Earth. 90 (1985) 12434–12444. https://doi.org/10.1029/JB090IB14P12434.

[80] D.E. Burch, D.A. Gryvnak, Optical and Infrared Properties of $Al_2O_3$ at Elevated Temperatures*, JOSA, Vol. 55, Issue 6, Pp. 625-629. 55 (1965) 625–629. https://doi.org/10.1364/JOSA.55.000625.

[81] J.M. Jones, P.E. Mason, A. Williams, A compilation of data on the radiant emissivity of some materials at high temperatures, J. Energy Inst. 92 (2019) 523–534. https://doi.org/10.1016/J.JOEI.2018.04.006.

[82] Q. Ren, H. Li, X. Wu, Z. Huo, O. Hai, F. Lin, Effect of the calcining temperatures of low-

grade bauxite on the mechanical property of mullite ceramics, Int. J. Appl. Ceram. Technol. 15 (2018) 554–562. https://doi.org/10.1111/ijac.12815.

[83] G. Buxbaum, G. Pfaff, Colored Pigments, in: Ind. Inorg. Pigment., Wiley, 2005: pp. 99–162. https://doi.org/10.1002/3527603735.

[84] T.A. Blake, T.J. Johnson, R.G. Tonkyn, B.M. Forland, T.L. Myers, C.S. Brauer, Y.-F. Su, B.E. Bernacki, L. Hanssen, G. Gonzalez, Methods for quantitative infrared directional-hemispherical and diffuse reflectance measurements using an FTIR and a commercial integrating sphere, Appl. Opt. 57 (2018) 432. https://doi.org/10.1364/AO.57.000432.

[85] Thermo Scientific, Product Specifications - Thermo Scientific Nicolet iS50 FTIR Spectrometer, (n.d.).

[86] Accuris Instruments, Analytical Balance W3100 Series - Operating Manual, (n.d.).